\documentclass{jfm}

\usepackage{graphicx}
\usepackage{newtxtext}
\usepackage{newtxmath}
\usepackage{xcolor}
\usepackage{natbib}
\usepackage{hyperref}
\hypersetup{
    colorlinks = true,
    urlcolor   = blue,
    citecolor  = black,
}

\newcommand{\RomanNumeralCaps}[1]

\newcommand{\hx}{\hat{x}}
\newcommand{\hy}{\hat{y}}
\newcommand{\rmd}{\mathrm{d}}
\newcommand{\br}{{\boldsymbol r}}
\newcommand{\reta}{\br_\eta}
\newcommand{\rz}{{\br}_{\barz}}
\newcommand{\pbz}{(\barz)}
\newcommand{\fV}{f^\mathrm{V}}
\newcommand{\fS}{f^\mathrm{S}}

\newcommand{\fVz}{\fV\pbz}
\newcommand{\fVeta}{\fV_{\eta}}
\newcommand{\fSeta}{\fS_{\eta}}
\newcommand{\fVetaz}{\fVeta\pbz}
\newcommand{\fSetaz}{\fSeta\pbz}

\newcommand{\rms}{_\mathrm{rms}}
\newcommand{\Taylor}{\lambda_\mathrm{T}}
\newcommand{\urms}{\tilde{u}}
\newcommand{\Linf}{L_\infty}
\newcommand{\Lb}{L_\beta}
\newcommand{\Ln}{L_\nu}
\newcommand{\Reinf}{Re_\infty}
\newcommand{\Rel}{Re_\lambda}
\newcommand{\pp}{\partial}
\newcommand{\order}[1]{\mathcal{O}(#1)}
\newcommand{\lamci}{\lambda_{ci}}
\newcommand{\Tinf}{T_\infty}

\newcommand{\bom}{{\boldsymbol \omega}}
\newcommand{\bu}{\mathbf{u}}
\newcommand{\burms}{\bu\rms}

\newcommand{\barz}{\bar{z}}

\usepackage{subfig}
\usepackage{comment}
\usepackage{amsfonts}
\usepackage{amsmath}
\usepackage{enumitem}
\newcommand{\overbar}[1]{\mkern 1.5mu\overline{\mkern-1.5mu#1\mkern-1.5mu}\mkern 1.5mu}
\usepackage[percent]{overpic}

\newcommand{\noshow}[1]{}

\shorttitle{Vortex structures under dimples and scars}
\shortauthor{J. R. Aarnes, O. Babiker, A. Xuan, L. Shen and S. Å Ellingsen}

\title{Vortex structures under dimples and scars in turbulent free-surface flows}

\author{Jørgen R. Aarnes\aff{1}\corresp{\email{jorgen.r.aarnes@ntnu.no}},
Omer Babiker\textsuperscript{1}, 
Anqing Xuan\textsuperscript{2}, 
Lian Shen\textsuperscript{2} 
\and Simen Å. Ellingsen\textsuperscript{1}}
\affiliation{\aff{1} Department of Energy and Process Engineering, Norwegian University of Science and Technology, Trondheim, Norway
\aff{2} Department of Mechanical Engineering and Saint Anthony Falls Laboratory, University of Minnesota, Minneapolis, Minnesota 55455, USA}

\begin{document}
\maketitle

\begin{abstract}
Turbulence beneath a free surface leaves characteristic long-lived signatures on the surface, such as upwelling `boils', near-circular `dimples' and elongated `scars', easily identifiable by eye, e.g., in riverine flows. In this paper, we use Direct Numerical Simulations to explore the connection between these surface signatures and the underlying vortical structures. We investigate dimples, known to be imprints of surface-attached vortices, and scars, which have yet to be extensively studied, by analysing the conditional probabilities that a point beneath a signature is within a vortex core as well as the inclination angles of sub-signature vorticity. The analysis shows that the likelihood of vortex presence beneath a dimple decreases from the surface down through the viscous and blockage layers in a near-Gaussian manner, influenced by the dimple’s size and the bulk turbulence. When expressed as a function of depth over the Taylor microscale $\lambda_T$, this probability is independent of Reynolds and Weber number. Conversely, the probability of finding a vortex beneath a scar increases sharply from the surface to a peak at the edge of the viscous layer, at a depth of approximately $\lambda_T/4$. Distributions of vortical orientation also show a clear pattern: a strong preference for vertical alignment below dimples and an equally strong preference for horizontal alignment below scars.
Our findings suggest that scars can be defined as imprints of horizontal vortices approximately a quarter of the Taylor microscale beneath the surface, analogous to how dimples can be defined as imprints of surface-attached vertical vortex tubes.

\end{abstract}


\section{Introduction}
Turbulence near free surfaces plays a key role in a multitude of environmental, biological, and engineering processes, e.g., in the fluxes of heat and gas across the air--water interface in rivers and oceans. Due to the difficulty of bulk flow measurements in the field, relating free-surface features to the sub-surface flow structures that cause them is of great practical importance \citep{muraroFreesurfaceBehaviourShallow2021}. Of particular interest is the very near surface region, which affects the gas transfer of the low diffusivity gasses that make up the majority of the atmosphere \citep{herlina2014}. A more detailed coupling between identifiable patterns on the free surfaces of lakes, rivers, and oceans to the near-surface flow dynamics could mean a breakthrough in using data collected from cheap remote sensing technology as input for, e.g., estimating greenhouse gas discharge from rivers, similar in magnitude to the total gas flux across the ocean surface, and to the total forest carbon uptake rate \citep[e.g.][]{brinkerhoff2022}.

\begin{figure}%
    \centering
    \includegraphics[width=\linewidth]{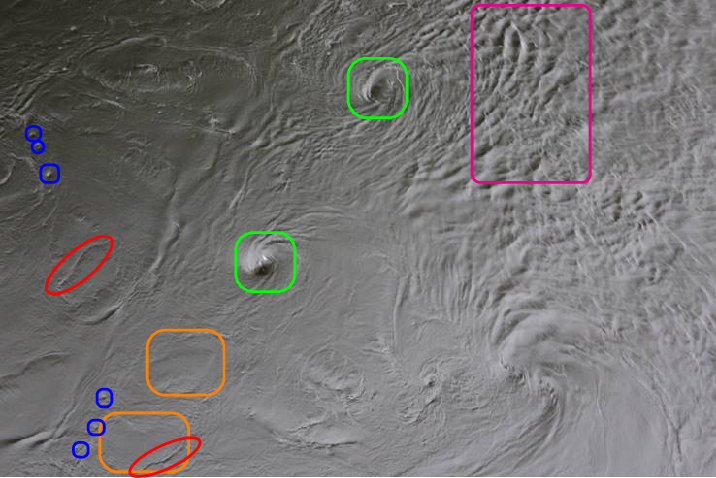}
    \caption{
    Snapshot of the river Nidelva in Trondheim, exhibiting a multitude of surface deformations. A small selection of dimples (blue), scars (red), upwelling boils (orange), and waves (magenta) are marked. The largest dimples (green) are von K\'arm\'an vortices shed from a nearby bridge pillar. Photo by Klervie le Bris.}%
    \label{fig:nidelva}%
\end{figure}

The free surface of a turbulent flow exhibits a multitude of deformations that manifest as identifiable surface patterns. Among them, we typically find upwellings, dimples, waves, scars, bubbles, break up of the surface, and droplet formations; a useful taxonomy was given by \citet{brocchiniDynamicsStrongTurbulence2001} and reviewed and applied by \citet{muraroFreesurfaceBehaviourShallow2021}. The present study focuses on continuous surfaces with no break-up, where gravity is dominant and surface tension is weak, the most common state in terrestrial bodies of water with flow \citep[see][§\,7]{brocchiniDynamicsStrongTurbulence2001}. Here, the turbulence has enough energy to disrupt the surface, without breaking it up, and the persistent surface deformations are mainly caused by intermittent upwelling and downwelling events, rushes of fluid towards and away from the surface (see figure \ref{fig:nidelva}).
At the surface, these give rise to regions where the surface divergence, $\beta = \partial_x u + \partial_y v = -\partial_z w$ when the surface is nearly flat and normal to $z$, is strongly positive (upwelling) or negative (downwelling). Experiments \citep{Rashidi1997,kumar98,Gakhar2022} and simulations \citep{nagaosa1999, khakpour12} show how these can be caused by turbulent structures ejected from the shear layer, or above a canopy \citep{mandel19}, that are advected towards the surface. Efforts to properly characterise upwelling boils go back a long time, notably by \cite{matthes1947}.

Dimples --- persisting near-circular indentations at the surface --- have long been observed at the edges of upwelling boils \citep[see][]{banerjee94, longuet-higgins96}. These dimples emerge when sub-surface vortices break up and attach to the surface, as observed in studies on vortex ring connection to free surfaces, both experimentally \citep{bernal1989,willert1997} and in simulations \citep{zhang1999, terrington2022}. Also appearing on the edges of upwelling regions are the scars, a type of surface pattern that has attracted less attention in the literature. These elongated, valley-like patterns on the surface mark regions where the fluid is accelerated from a region of strong positive divergence (upwelling) to a region of strong negative divergence (downwelling), causing a local pressure deficit and a concomitant depression of the surface. Both dimples and scars share several practically important traits: they are both local, persistent surface depressions imprinted by strong coherent turbulent structures beneath, they are readily detectable on the surface by eye or computer vision, and both are closely connected to upwelling and downwelling events which are the main drivers of surface renewal facilitating surface fluxes \citep{kermani09}. This motivates the main focus of the present paper: To increase the understanding of dimple and scar patterns by quantifying their sub-surface patterns in turbulent free-surface flows.

The interest in quantifying the turbulence near the surface for the estimation of heat and gas transfer has inspired new techniques for estimating turbulence properties close to the surface. \emph{In situ} measurements which penetrate the surface are routinely employed and can be used to quantify the surface and to make statistical inferences about surface properties or the turbulence below. However, these can only be measured at single points or along trajectories, and upscaling to cover many sites over large areas is forbiddingly costly. An attractive alternative is noncontact measurements from above which use only information from the surface itself; see the the recent overview by \cite{Dolcetti2022}. It requires knowledge of what information about the fluid flow the observed surface features can provide. A step in this direction is taken by \cite{babiker2023}, who detect and track the dimples at the surface by applying wavelet analysis to surface elevation data (from direct numerical simulations). The detected dimples show a strong correlation to variations in the surface divergence, $\beta$, and may therefore act as a proxy for measuring the effects of upwelling events on surface renewal \citep{babiker2023} --- 
a beneficial connection as the surface divergence is linked to the mass transfer velocities of low solubility gasses and is often used in such models \citep[see, e.g.,][]{banerjee04, turney2013,kermani11}. 
A further step of statistical inferences from surface observations is to use machine learning. \cite{xuan23} explore how neural networks may be used to reconstruct the entire flow field below a free surface. Such data-driven methods have the potential to greatly improve remote sensing campaigns by reconstructing the turbulence or turbulent statistics with limited measurements but are still in their infancy for this application.

Of greatest practical interest is the surface boundary layer where turbulent mixing and surface renewal take place. 
A highly fruitful approach was pioneered by \cite{hunt1978} who applied a rapid-distortion model to analyse a turbulent boundary layer atop a wall moving with the mean flow velocity, and found two regions in the surface boundary layer dominated by different mechanisms: An inner, viscous region and an outer, source (or blockage) region. Although there are important differences in boundary conditions between a rigid surface and a deformable, free-slip interface, Hunt \& Graham's subdivision of the boundary layer conceptually applies also in free-surface turbulence. Hunt \& Graham's model has been expanded upon in later studies \citep{hunt84, teixeira2000, magnaudet2003}, and widely discussed in the literature on the surface boundary layer \citep[e.g.,][]{brumley1987, shenSurfaceLayerFreesurface1999,shen2000, nagaosa03, flores2017}. In short, the two regions have been tied to the two boundary conditions at the surface: the inner region, known as the viscous (sub)layer, to the dynamic boundary condition, and the outer region, the blockage layer, to the kinematic condition. We return to this in \S\,\ref{subsec:flowProp}. For now, we are content with stating that it is particularly advantageous to link surface observations to the different regions of the boundary layer of the free surface, as this yields a measure of the relevant length scales at which the effects act and what physical mechanisms can be tied to the observations.

The present study aims to elucidate how and to what extent dimples and scars are coupled to the sub-surface dynamics of the flow. When exploring the nature of scars, we immediately run into the fundamental question of what a scar really is. Whereas a number of authors offer qualitative descriptions, at present we know of no precise set of criteria. To aid us we propose the following conceptual framework. 

\subsection{Concepts for surface-feature identification}

In order to establish a firmer connection between surface features and the sub-surface structures they are inextricably linked to, we will introduce two concepts pertaining to a class of surface features (the classes considered here are dimples and scars): the \emph{pseudo-causal definition} and a set of \emph{surface-only criteria}. When a well-matched pair of these has been identified, it prepares the ground for matching surface imprint observations to sub-surface flow phenomena in a quantitative way. For dimples, both of these can be formulated based on current knowledge while it is one of our goals herein to propose these for the case of scars. 

A pseudo-causal definition (PCD) of a class of surface features states that a `true' member of said class is the surface imprint of a specified sub-surface structure. It should be formulated such that an observed feature is unambiguously sorted into at most one class. For the case of dimples, the appropriate PCD is that a `true' dimple is the imprint of a surface-attached vortex and appears where said vortex is attached. For practical use, for instance, in direct numerical simulation (DNS) data, a quantitative formulation is required, for example by identifying a vortex using the $\lambda_2$ criterion which we will discuss in detail later. Strictly speaking, the causal narrative that one typically uses (e.g., that the dimple is `caused by' the presence of the vortex) is not correct: the two appear together, simultaneously, and are two expressions of one and the same flow event (note that propagating surface waves are different in this respect). Hence, the label `pseudo-causal'. The PCD is of limited practical use in a field setting because detailed knowledge of the flow is necessary. On the other hand, it expresses a physical understanding of the flow picture. In practice one can hardly achieve the ideal level of universality and user independence: for instance, the $\lambda_2$ criterion for vortex identification requires a user-defined threshold value, and a given surface feature is the result of the full nearby flow field, not merely the single most dominant structure. The very concept of dividing (material or Eulerian) sub-volumes of a continuous flow into `vortex' and `not vortex' is inherently ambiguous and non-unique. A level of pragmatism is inescapable also in formulating a PCD, and such is the nature of the problem at hand. 

A set of surface-only criteria (SOC) of a class of surface features, refers only to the free surface using no information about sub-surface flow. It is heuristic, pragmatic and might be context-dependent, yet should be formulated so that it can also be quantitatively applied to decide whether or not a point on the surface is part of a `dimple' or `scar'. When speaking of scars and dimples in the following, we refer to the SOC, i.e., only based on surface appearance, in contrast to `true dimples' and `true scars' as decided by the PCD. A suitable SOC for dimples was demonstrated by \citet{babiker2023} (without calling it such): a dimple is a surface indentation which persists for a minimum time and is sufficiently circular. The SOC should be usable in practical field settings where only the surface is observed, but is purely descriptive and expresses no understanding of the underlying flow.

Once a successful PCD-SOC pair has been identified, it paves the way for inferring quantitative information about sub-surface flow from surface observations in a structured way, for instance, via statistical investigations, such as we present herein. 
A successful PCD-SOC pair must have a high level of agreement. For dimples, \citet{babiker2023} showed that the mentioned PCD and SOC when applied to DNS data produced the same identification for practical purposes. Since the dynamics of surface-attached `bathtub' vortices are comparatively well understood, the success of this PCD-SOC pair is not particularly surprising.

\subsection{Procedure and outline}

We explore how dimples and scars are linked to sub-surface structures by considering the conditional probability that the points directly beneath dimples and scars lie inside a vortex core. Furthermore, we consider the statistics of how the vorticity beneath these surface features is inclined, tying vortex tubes below dimples and scars to two different preferred directions for the vortex inclination. 

The dimples in our analysis are identified by building upon the knowledge of their PCD-SOC pair. As the equivalent PCD-SOC pair is unknown for scars, we proceed in an explorative manner in scar detection, by suggesting what amounts to a candidate SOC for scars and then studying statistically the flow field beneath the resulting structures. Based on the statistical results we are able to identify and argue for a successful formal classification of scars in the form of a PCD-SOC pair.
The data analysed stems from DNS for a selection of Reynolds and Weber numbers.

The paper is organised as follows: In \S\,2 we provide details on the mathematical formulation of our flow and outline key aspects of the DNS. In \S\,3 the DNS results are presented and analysed. \S\,4 contains our main analysis of vortices and scars in the flow before conclusions are drawn in \S\,5.

\section{Flow problem and DNS}

The flow problem analysed is isotropic turbulence interacting with a free surface. The problem is that of \cite{guoInteractionDeformableFree2010}, but in the present study, several cases with different Reynolds numbers and Weber numbers are considered, which also affects the choice of grid resolutions and the turbulent properties. Details on the flow problem, simulations, and grid independence can be found in \cite{guoGenerationMaintenanceWaves2009,guoInteractionDeformableFree2010} and \cite{xuanConservativeSchemeSimulation2019,xuan2022}. Only a brief synopsis is given here.

\subsection{Flow problem}
The governing equations for the flow are the incompressible Navier--Stokes equations and the continuity equation:
\begin{align}
    \frac{\partial u_i}{\partial t} + \frac{\partial (u_i u_j)}{\partial x_j} & = - \frac{\partial p}{\partial x_i} + \frac{1}{\Rey} \frac{\partial ^2 u_i}{\partial x_j \partial x_j} + f_i \, , \label{eq:ns_mom}\\
    \frac{\partial u_i}{\partial x_i} & = 0 \, , \label{eq:ns_conti}
\end{align}
where $u_i$ ($i = 1, 2, 3)$ are the instantaneous velocity components, $p$ is the dynamic pressure normalised by $\rho U^2$, where $\rho$ is the density and $U$ is the characteristic velocity scale. The term $f_i$ on the right-hand side is a depth-dependent forcing term (defined below). The Reynolds number is $Re = U L / \nu$, with characteristic length $L$ and kinematic viscosity $\nu$. 

The flow variables are computed in a three-dimensional domain that is periodic in the horizontal directions ($x_1, x_2$, i.e., $x, y$) and restricted above by a deformable free surface at $x_3 = z = -\eta$, where $\eta = \eta(x,y,t)$ is the surface elevation and positive $z$-direction is taken to be from the surface and down into the flow (with average depth $\Bar{z} = 0$ at the surface). At the bottom boundary, a free-slip boundary condition is used to eliminate the viscous effects of the bottom boundary layer.

The flow is subjected to the kinematic and dynamic boundary conditions at the free surface. The kinematic boundary condition states that the upper boundary is material, so that particles on the surface remain on the surface from a Lagrangian point of view, that is, we can have no penetration of the free surface. The dynamic boundary condition states the stress balance at the free surface. To ensure that the surface is well resolved, an undulating grid that adheres to the free surface is used. \citep[For details on the computational methods, boundary resolutions and algebraic transformation used in the grid transformation, see][]{xuanConservativeSchemeSimulation2019}.

Turbulence is generated in the vertical centre region of the domain and is transported towards the surface. The turbulence is generated by a  random linear forcing~\citep{lundgren2003,rosalesLinearForcingNumerical2005} with forcing function $f_i = a_0 F(z) u_i$~\citep{guoGenerationMaintenanceWaves2009}, which is proportional to the instantaneous velocity $u_i$, a forcing parameter $a_0$ ($a_0^{-1}$ approximately determines a turnover time of the forced turbulence and is selected such that $U = a_0 L$), and the depth-dependent function $F(z)$. $F(z)$ is 1 in the centre region of the domain, falls off through a damping region, and is zero in the region denoted the `free region', which covers the depth $z \in [-\eta, 0.5\pi L]$. The function $F(z)$ is designed in such a way to ensure that the imposed forcing does not directly affect the flow in the vicinity of the surface, i.e., in the free region \citep[details in][]{guoGenerationMaintenanceWaves2009}. 

The total domain size is $L_x \times L_y \times L_z = 2\pi L \times 2\pi L \times 5\pi L$, with the vertical direction spanning the largest length due to the requirements of the forcing function to generate turbulence that acts as decaying isotropic turbulence as it enters the free region. Only results from the free region are used in the analysis. The grid used in the simulations has $256\times256\times660$ points for the high Reynolds number cases and $128\times128\times348$ points for the low Reynolds number cases (case details in  \S\,\ref{subsec:flowProp}). The grid is equidistant in $x$- and $y$-direction and stretched in $z$-direction to ensure a very fine grid near the free surface. The stretching enhances resolution of the turbulence dynamic processes near the surface. The duration of the simulations are $160 T_\infty$--$200 \Tinf$ (where $\Tinf$ is the eddy turnover time) after becoming statistically stationary. Sampling is done every at intervals of $0.0127\Tinf$--$0.0182\Tinf$.

The governing equations are discretised on the computational grid using a pseudo-spectral method in $x$- and $y$-directions and by a finite-different method in the $z$-direction. This hybrid discretisation leverages the strengths of each method, with the pseudo-spectral method providing high accuracy in horizontal planes and the finite-difference approach handling the grid clustering flexibly. To accurately model the interactions between near-surface turbulence and free-surface motions, the simulations are performed on a deforming curvilinear grid that fits the free surface. The free-surface elevation is updated by using a second-order Runge--Kutta method to integrate the kinematic boundary condition. Within each stage of the Runge--Kutta integration, the Navier--Stokes equations (\ref{eq:ns_mom}--\ref{eq:ns_conti}) are solved using a fractional-step method to maintain the incompressibility constraint. The pressure Poisson equation, required by the projection step in the fractional-step method, involves variable coefficients arising from the grid transformation and is solved by a fixed-point iteration method \citep{xuanConservativeSchemeSimulation2019}. Using the curvilinear grid, despite its high computational demand, ensures that the complex dynamics within the free-surface turbulence boundary layer are accurately captured.

\subsection{Flow properties: length scales and nondimensional groups} \label{subsec:flowProp}

In this section we review and define the pertinent nondimensional groups and lengthscales we will later use in the analysis of turbulent flow beneath a free surface.
Three dimensionless numbers characterise the free-surface flow: The Reynolds number, the Weber number, and the Froude number. We have performed simulations for two Reynolds numbers, $Re = 2500$ and $Re = 1000$. In all simulations, the characteristic length and velocity scales are unity, while different values of viscosity are set to get different Reynolds number cases. To take into consideration the effects of surface tension, $\sigma$, we have simulated flow with two Weber numbers, $We = 20$ and $We = \infty$. The Weber number is defined as $We = \rho U^2 L / \sigma$, hence $We = \infty$ is the theoretical case of no surface tension. The Froude number, defined by $Fr = U/\sqrt{g L}$, where $g$ is the gravitational acceleration, is fixed at 0.1 for all simulations. As $We \gg 1$ and $Fr \ll 1$, all simulations are performed for flows with gravity-dominated surface motion.
 
In the absence of mean flow, the turbulent Reynolds number is an appropriate measure. It is defined as $\Reinf = {\urms l} / {\nu}$, where $\urms$ is the `representative velocity' \citep[][Section  3.1]{tennekes1972first}, defined as $\urms^2 = \frac{1}{3} \overbar{u_iu_i}$, and $l$ is the macroscale of the turbulence. Thus, $\urms^2$ equals two-thirds of the turbulent kinetic energy. The exact manner of averaging implied by the overbar is important because most statistical turbulence properties beneath a free surface vary with depth, a point we will discuss below.

While the root-mean-square velocity is computed directly from flow variables, the macroscale is approximated (under the assumption of isotropic turbulence, verified in \S\,3.1) by  $l \approx {\Taylor} \Rel / 15 = \urms^3/\epsilon$, where $\Rel = \urms \Taylor / \nu$ is the Taylor Reynolds number, and $\Taylor = \urms\sqrt{15 \nu / \epsilon}$ is the Taylor microscale \citep[][p.~67]{tennekes1972first}. To compute the latter, the dissipation $\epsilon = 2 \nu \overbar{s_{ij}s_{ji}}$ is computed from the stress components $s_{ij}=(\pp_iu_j+\pp_ju_i)/2$
whence the Kolmogorov length scale, $L_K = (\nu^3 / \epsilon)^{1/4}$, can be calculated.
A more widely used length scale than the macroscale is the integral scale of the turbulence, approximated by $\Linf = l / 2$ \cite[p.~273]{tennekes1972first}, whence we arrive at the form of turbulent Reynolds number most commonly found in literature on free-surface flows: $\Reinf = 2 \urms\Linf / \nu$. The integral length scale and an appropriate velocity scale yields the eddy turnover time, $\Tinf$. Here,  $\Tinf = \Linf / \urms$.

The presence of a free surface fundamentally changes the properties of the turbulence immediately beneath it. Nearest the surface a thin \emph{viscous surface layer} is formed, beneath which is found a thicker \emph{blockage layer} \citep[see e.g.,][]{shenSurfaceLayerFreesurface1999}. 
The viscous layer has thickness $\Ln\sim \Reinf^{-1/2}$ and occurs because the tangential viscous stress at the free surface must be zero, and the horizontal vorticity must also be small. Within the viscous layer, the horizontal vorticity fluctuations reduce rapidly towards the surface. The blockage layer extends deeper, to a depth $\Lb\sim \order{\Linf}$, and occurs because velocities must be tangential to the surface. Thus horizontally oriented vortex tubes must either move away from the surface or break up and attach to it. Based on measurements by \cite{brumley1987} the viscous surface layer thickness is conventionally taken to be $L_\nu = 2 \mathrm{Re}_\infty^{-1/2} L_\infty$ \citep[see discussion in][]{calmetStatisticalStructureHighReynoldsnumber2003}. No single definition for $\Lb$ has become standard, yet $\Lb=\Linf$ is often used for convenience \citep[see e.g.,][]{magnaudet2003, guoInteractionDeformableFree2010,herlina2014}. A more detailed discussion of the layered structure of free-surface flow may be found, e.g., in \citet{shenSurfaceLayerFreesurface1999}.

We note in passing that the literature is not consistent regarding the factor $2$ in the expression for $\Ln$. The reason seems to be that both the turbulent macroscale and the integral scale are in common use, which are only a factor $2$ apart and are occasionally conflated. 

In the present study, nondimensional groups and characteristic lengthscales which characterise the turbulence are calculated in the bulk. Recall that we consider only the free region in our analysis. Since there is no forcing in this region, the turbulence decays as it approaches the surface, hence, dissipation, root-mean-square velocity, etc., vary with depth (details in \S\,\ref{sec:flowFeatures}). Thus, we have no single measure of bulk properties. To make a comparison between the different cases as clear as possible a measure of bulk properties is made at a single reference depth, where averages and root-mean-square values (implied by the overbars in the definitions for $\urms$ and $\epsilon$) are computed plane-wise, before being averaged in time to get better statistics. The selected reference depth is outside the blockage layer to ensure that the theory for isotropic turbulence is a reasonable approximation, yet close enough to the blockage-layer boundary to minimise turbulence decay between the reference plane and the near-surface region. In all cases, the reference depth is chosen as $\pi/3$ ($ \approx 1.6\Linf$ for case 1).

A summary of the most important flow properties can be found in table \ref{tab:flowProp}.

\begin{table}
    \begin{center}
    \begin{tabular}{c c c c c c c c c c c}
    Case &  Re  & We & Fr   & $\mathrm{Re}_\infty$   & $\mathrm{Re}_\lambda$ &$L_\infty$     & $\lambda_T$ & $L_K$ & $L_\nu$\\ 
    1  & 2500  & $\infty$  & 0.1   &  433  & 91 & 0.64 & 0.24 & 0.0135 & 0.061 \\ 
    2  & 2500  & 20        & 0.1   &  368  & 74 & 0.57 & 0.23 & 0.0135 & 0.059 \\ 
    3  & 1000  & $\infty$  & 0.1   &  131  & 44 & 0.54 & 0.36 & 0.028  & 0.094 \\ 
    4  & 1000  & 20        & 0.1   &  164  & 50 & 0.61 & 0.37 & 0.027  & 0.096 \\ 
    \end{tabular}
    \caption{Flow and turbulence properties. From left: case number, Reynolds number, Weber number, Froude number, turbulent Reynolds number, Taylor Reynolds number, integral length scale, Taylor length scale, Kolmogorov length scale, viscous layer thickness.}
    \label{tab:flowProp}
    \end{center}
\end{table}


\section{Flow in the near-surface region} \label{sec:flowFeatures}

Before discussing surface manifestations, we discuss some general properties of turbulent flow beneath a free surface, in particular the appearance of the blockage and viscous layers. While in a large part known, the concepts and general properties in \S\,\ref{sec:flowFeatures} provide the backdrop against which further analysis can be compared.

\subsection{Surface effects on the velocity and vorticity components} \label{subsec:velvort}

\begin{figure}%
    \centering
    \includegraphics[width=1\linewidth]{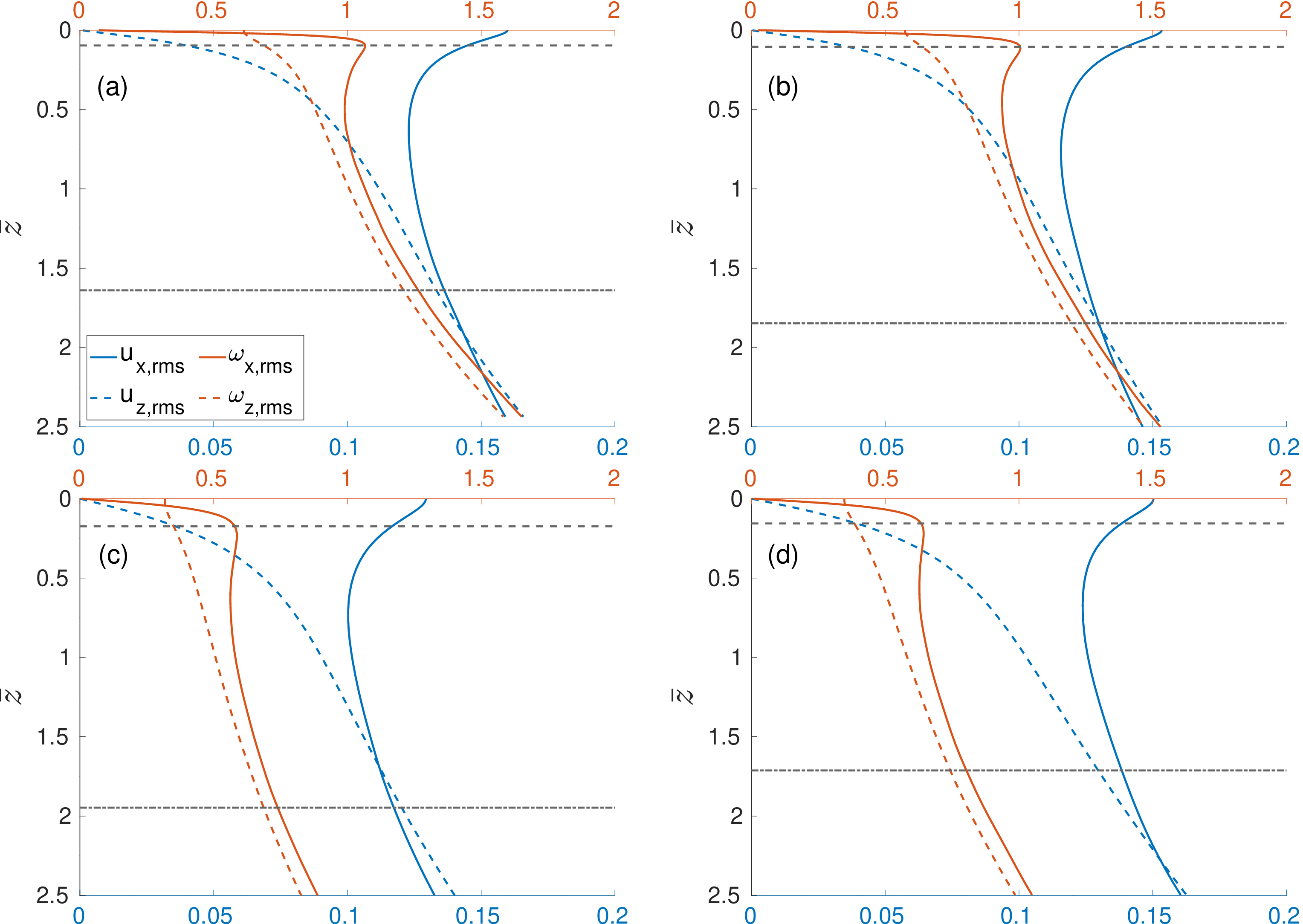}%
    \caption{Root-mean-square values of velocity (blue curves and bottom-of-panel abscissae) and vorticity (red curves, top-of-panel abscissae) for cases 1--4 in figures (a)-(d), respectively. Flow variables plotted against averaged depth, $\Bar{z}$, normalized by the integral length scale ($L_\infty$). The horizontal lines mark the viscous boundary layer limit (dashed) and the reference depth (dash-dotted).
    }%
    \label{fig:velvortRMS}%
\end{figure}

Consider components of the velocity vector $\burms$ and the vorticity field $\bom\rms=\left(\nabla\times \bu\right)\rms$ as a function of depth, for cases 1--4 in figure \ref{fig:velvortRMS}(a--d), respectively; the subscript `rms' denotes root-mean-square, i.e., $(\cdots)\rms=[\overbar{(\cdots)^2}]^{1/2}$ taking the average over $x,y,t$, but not $\barz$. Note that only a single horizontal component is included as the flow is horizontally isotropic.

An average measure for `horizontal' grid plane depth is used here and later, denoted $\barz$. In short, the zero-level of $\bar{z}$ is the surface elevation measurement, i.e., $z(x,y,\barz)|_{\barz=0}=-\eta$, 
and $\bar{z} > 0$ is the average distance from grid points $(x_i,y_j,z_k)$ on grid plane $k>0$ to the surface points $(x_i,y_j,z_0)$. In effect, $\barz$ is used to approximate the vertically undulating grid as a vertically stretched Cartesian grid in the analysis.
This convention enables straightforward comparison of flow variables at a grid plane on the undulating grid without interpolation. As the variations in surface elevation in a grid plane are small ($\mathcal{O}(10^{-4} L_\infty)$ for all cases), this is an acceptable convention. 
\citep[An alternative to $\barz$ is to use a signed distance function for the depth, which necessitates interpolation for all measurement values, see][]{guoInteractionDeformableFree2010}.

For all cases, we observe the redistribution of kinetic energy from the vertical to the horizontal velocity components as the surface is approached from below. Here the vertical component tends towards zero, as reflected in the kinematic boundary condition which restricts movement in the vertical direction for low Froude number flows with small surface deformation. Further, we recognize that the flow is roughly isotropic for $\bar{z} \gtrsim 1.5L_\infty$, agreeing with the notion that the blockage layer thickness is of order $\mathcal{O}(L_\infty)$. It also suggests that the selected depth for bulk measurements, $\pi/3$ ($\approx 1.6 L_\infty$ for case 1) is appropriate, as it is outside the region where surface effects influence the flow.

The developments of the vorticity as the surface is approached demonstrate the influence of the dynamic boundary condition. Due to the zero-stress condition at the surface, the horizontal vorticity falls sharply through the viscous boundary layer while only a small effect can be seen for the vertical vorticity. The viscous boundary layer thickness, $\Ln$ defined in \S\,\ref{subsec:flowProp}, coincides well with the local maximum in horizontal vorticity for all cases, although the local vorticity maxima are less sharp for the lower Reynolds number cases. This supports the notion that the viscous sublayer thickness can be approximated by the depth of the shallowest maximum of the horizontal root-mean-square vorticity components \citep[see, e.g.][]{calmetStatisticalStructureHighReynoldsnumber2003}. 

Observe that the vertical velocity component decreases steadily through the blockage layer, from the isotropic bulk flow to nearly zero at the surface as dictated by the kinematic boundary condition. This is the blockage effect which forces the turbulence to become increasingly anisotropic. The onset of blockage in the outer part of the blockage layer is gradual, increasing as the surface is approached.

\subsection{Distribution of vortex inclination angles through the surface boundary layer}

The free-surface motions in the flow examined in this study are small, due to the low Froude number. Consequently, the dynamic boundary condition of zero tangential shear stress corresponds to small horizontal vorticity, making $\omega_z$ the only significant vorticity component at the surface. 
For a more detailed picture of the vortices in the vicinity of the surface, we quantify the orientation of the vortices in the flow by computing the distribution of the inclination angle of the vorticity vector along horizontal planes.
\citet{shenSurfaceLayerFreesurface1999} reported vorticity orientation in the presence of a shear flow. The absence of a mean flow in our case allows us to isolate particular features related to the turbulent vortices in the vicinity of the free surface. 

The inclination angle of $\bom$ is measured in a horizontal-vertical plane. Any orientation is equivalent, so we arbitrarily choose the $zx$-plane. The angle is computed by:
\begin{align}\label{eq:thxz}
    \theta_{zx} = \arctan(\omega_x / \omega_z) \, .
\end{align}
We consider statistics of the inclination angle weighted by the normalised magnitude of the $zx$-projection of $\bom$, similar to \cite{moin1985}, i.e., we compute the statistics of $\kappa_{zx}\theta_{zx}$, where the weighting factor is computed for each grid point by:
\begin{align}\label{eq:omxz}
    \kappa_{zx} = \frac{\omega_x^2 + \omega_z^2}{\left\langle \omega_x^2 + \omega_z^2 \right\rangle}
\end{align}
where $\left\langle \cdot\right\rangle=(L_xL_y)^{-1}\int (\cdot)\,\rmd x\rmd y$ is the horizontal-plane average (depending on $t$ and $\barz$). 
Recall that the flow is horizontally isotropic, so the choice of the $zx$ plane in \eqref{eq:thxz} rather than any other vertical plane is arbitrary.

\begin{figure}%
    \centering
    \includegraphics[width=1\linewidth]{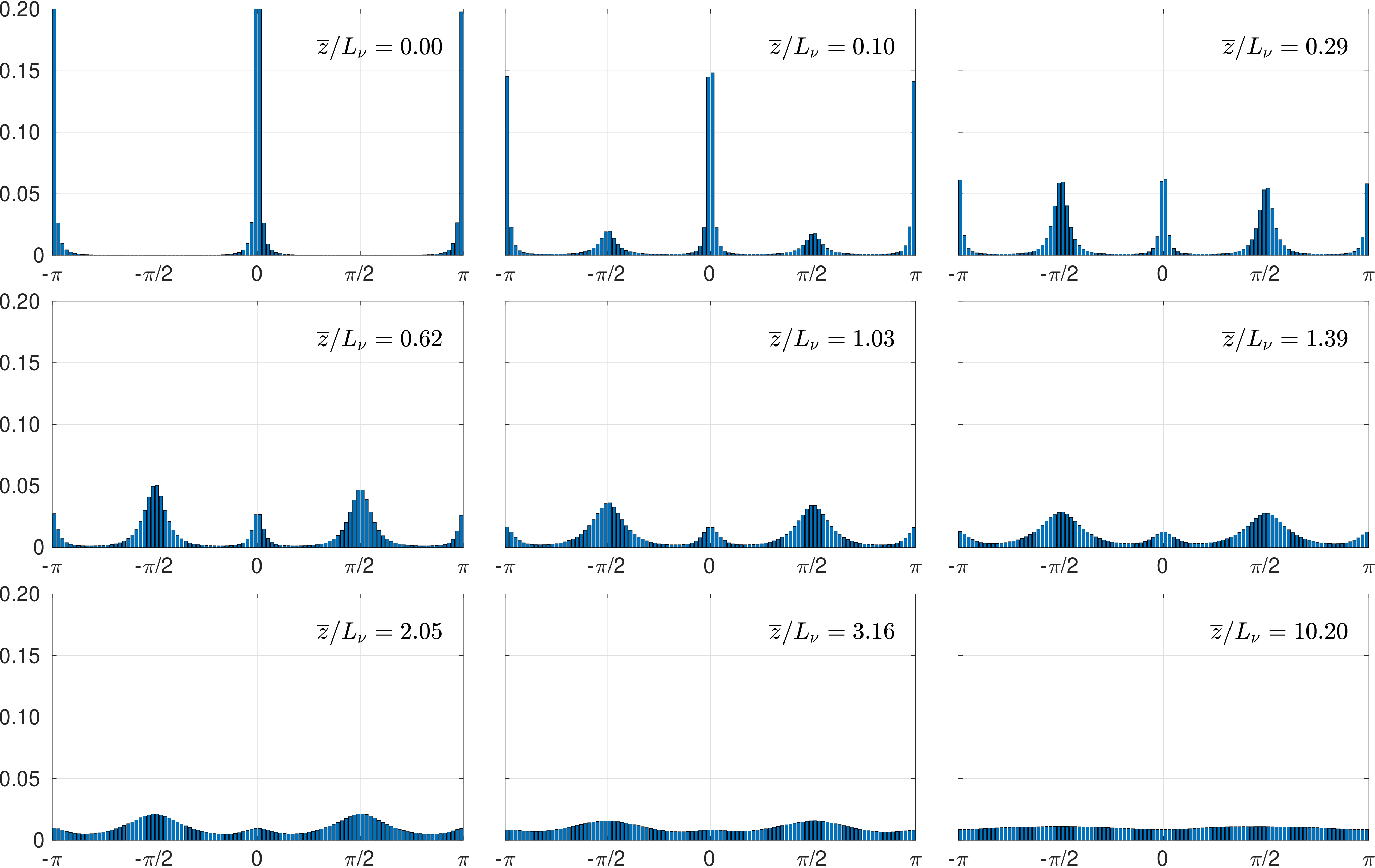}%
    \caption{Normalised histograms of weighted inclination angles $\theta_{zx}$ at different depths $\barz$ beneath the mean surface level for all $x,y,t$, in multiples of the viscous surface layer thickness $\Ln$.
   }%
    \label{fig:inclAngl}
\end{figure}

The distributions shown as histograms in figure \ref{fig:inclAngl} are results for case 1, computed for all time steps and all grid points in each grid plane at distances $\barz$ beneath the mean surface level (similar figures for the other cases can be found in Supplementary Material).
Nine values of $\barz$ are chosen, starting at the surface and extending throughout the blocking layer to approximately one integral length scale into the flow (recall from table \ref{tab:flowProp} that $\Linf \approx 10.5\Ln$ for case 1). Except near the base of the blockage layer where the flow approaches isotropy, $\theta_{zx}$ shows preferred distribution around $-\pi$, $-\pi/2$, $0$, $\pi/2$ or $\pi$, indicating that $\omega_z$ is either significantly smaller than $\omega_x$ ($\theta_{zx} = \pm \pi/2$) or larger than $\omega_x$ ($\theta_{zx}=0,\pm \pi$). Given that this probability distribution applies to vertical planes of any orientation, a high probability around $\theta_{zx} = \pm \pi/2$ indicates a predominance of horizontal vorticity, whereas a high probability around $\theta_{zx}=0,\pm \pi$ indicates the dominance of vertical vorticity.
We discern a gradual transition from vertical ($\theta_{zx} = 0, \pm \pi$) to horizontal ($\theta_{zx} = \pm \pi/2$) preferred orientation as one moves away from the surface. 
The dynamic boundary condition forces all vorticity to be normal to the free surface at the surface, hence the vertical orientation is strongly dominant there. Yet already in the middle of the viscous layer, the horizontal orientation is equally prominent, and it remains dominant in the upper part of the blocking layer before an essentially isotropic state is reached at $\barz\approx 10\Ln\approx \Linf$, roughly the deep boundary of the blocking layer. The results extend (and agree with) those of \cite{shenSurfaceLayerFreesurface1999} wherein only the surface and bulk were considered. Much light is shed on the transition from vertical to horizontal preference when dimples and scars are next considered.


\section{Dimples, scars, and the vortex dynamics in the surface boundary layer}

The results in \S\,\ref{sec:flowFeatures} give a general overview of sub-surface flow properties for the turbulent flow. Our principal goal is to understand the dynamics of the flow beneath vortices and, in particular, scars, including (but not limited to) the identification of a SOC-PCD pair for scars. To this end, we investigate statistical properties of the flow conditional on being beneath scars as identified from a heuristically proposed SOC. We begin, however, by performing for comparison the same procedure for the far better understood case of dimples, whence interesting insights also emerge.

In general discussions of scars and dimples in \S\S\,\ref{subsec:vortProb} to \ref{sec:probScars} we consider only case 1, for which the trends and observations are most clearly seen. In \S\,\ref{sec:ReWe_effects}, we compare these observations to the other cases in table \ref{tab:flowProp}. 

\newcommand{\lamth}{\lambda_{2,\mathrm{th}}}
\newcommand{\OmT}{\Omega_\mathrm{T}}

\subsection{Identification of dimples and vortices}\label{sec:dimples}

It was shown by \citet{babiker2023} that a suitable set of surface-only criteria (SOC) for dimples is (qualitatively speaking) that they are long-lived and approximately circular surface indentations, which complements the established understanding --- a pseudo-causal definition (PCD) in our terminology --- that a true dimple is the imprint of a surface attached vortex situated where the vortex tube terminates. With a simple computer vision method the SOC was implemented, showing excellent accuracy and precision in identifying true dimples according to the PCD, the latter implemented using the $\lambda_2$ method for vortex identification. Because the overlap is so good, it makes no practical difference for our statistical analysis of sub-dimple flow whether dimples are identified from the surface (SOC) or interior (PCD), so since the connection is not in question here we choose to use the latter henceforth. (This contrasts the later study of scars.)

The $\lambda_2$ method uses the three eigenvalues of the matrix $s_{ik}s_{kj} + w_{ik}w_{kj}$, where $w_{ij} = (\pp_iu_j - \pp_ju_i) / 2$ is the vorticity tensor and the eigenvalues $\lambda_i$ are ordered such that $\lambda_1<\lambda_2<\lambda_3$. Fluid points which satisfy $\lambda_2<\lamth\leq0$ are identified as belonging to a vortex core, where $\lamth$ is a threshold value which must be chosen carefully as it influences all our statistical analysis. After some consideration, we choose $\lamth=8/\OmT$ where $\OmT \equiv \urms/\Taylor$ is a characteristic frequency. (An equivalent relation is $\OmT^2=2\overline{s_{ij}s_{ji}}/15$.) 
Details are found in Appendix \ref{app:lam2threshold} where we also compare with an alternative vortex identification method.

\subsection{Detection and preliminary identification of scars} 
\label{sec:scarIdent}

Formulating a surface-only criterion for identifying scars is not equally obvious. 
The concept of scars simply signifies visual patterns on the surface that, according to \citet{brocchiniDynamicsStrongTurbulence2001}, `occur where flow on at least one side is downward causing a trough in the surface'. 
They described scars as cusp- or corner-like indentations, which appear as lines on the surface at the edge of upwelling boils and in regions adjacent to submerged horizontal vortices. Other studies have shown how pairs of counter-rotating vortices generate scars at the free surface: \cite{sarpkaya91} demonstrated this experimentally, generating the counter-rotating vortex pair through a submerged slit \cite[see also][]{sarpkaya96}. \cite{ohring91} explored the same phenomenon in simulations of laminar flow, focusing on how different Froude and Weber numbers affect the position of the scar relative to the submerged vortices.

\begin{figure}%
    \centering
    \includegraphics[width=\linewidth]{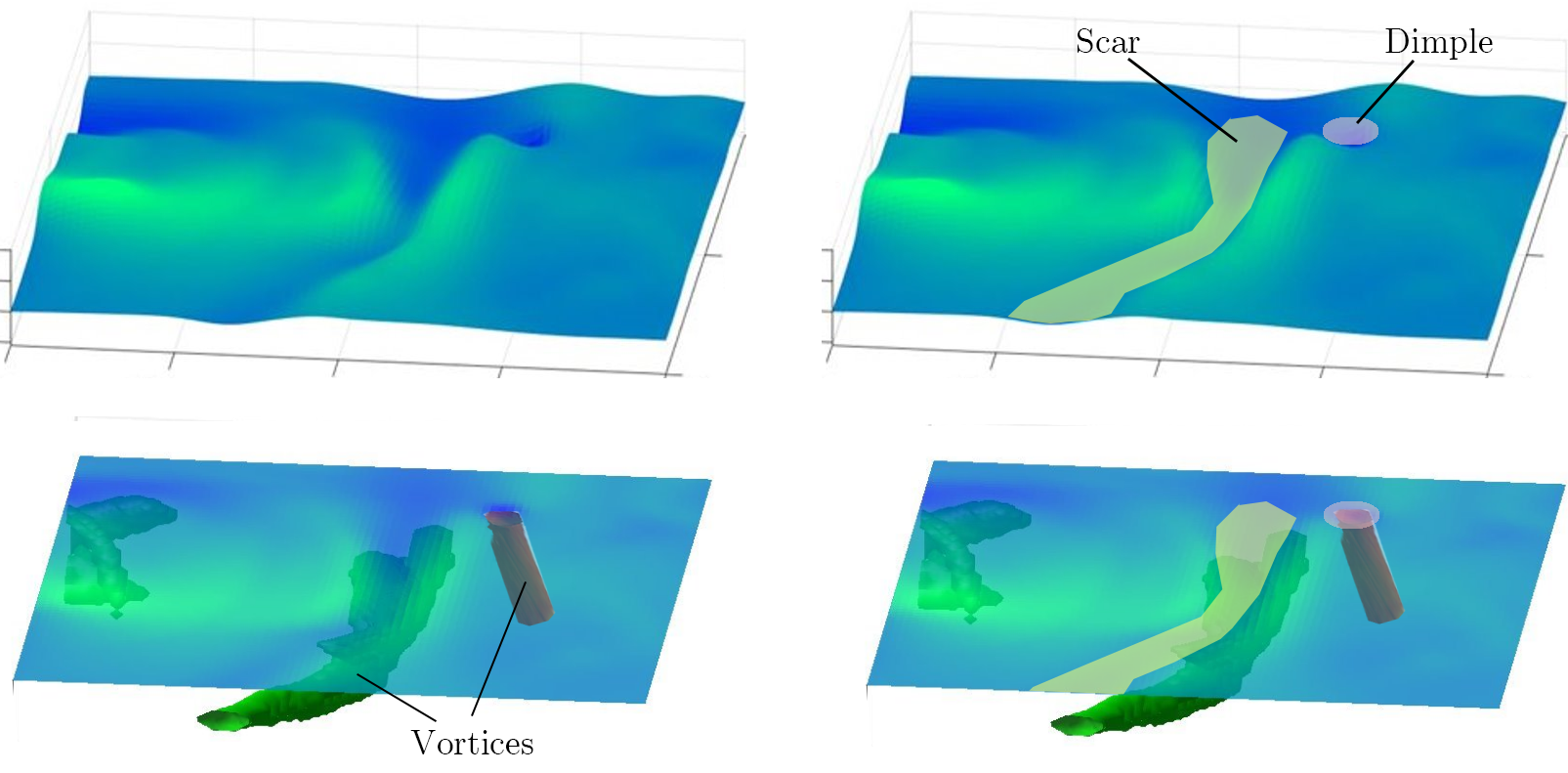}
    \caption{
    Conceptual sketch showing a representative situation, to illustrate the concepts of dimples, scars and vortices and how they typically occur together. Surface elevation (a),  together with detected surface features (b), sub-surface vortices (c), and both (d). Vortices are illustrated as isosurfaces of $\lambda_2$ in the sub-surface velocity field, coloured green and red for distinguishability, and some small vortices have been removed for visibility in (c) and (d).}%
    \label{fig:scar}%
\end{figure}

In the present study, we observe submerged horizontal vortices at the edges of upwelling events, with a scar located directly above such a vortex (illustrated in figure \ref{fig:scar}). This is similar to the low Froude number cases shown in \cite{ohring91}, where they saw a scar-like indentation directly above the vortex (as opposed to higher Froude numbers where the scars appear to the side of the vortex). Hence, we find it reasonable to hypothesize that scars are inextricably linked to the said vortex, so that scars, like dimples, are coupled to a specific sub-surface coherent turbulent structure which is large and long-lived. Such information can be used both to make inferences about sub-surface behaviour below the scars detected at the surface, and, reversely, to classify scars as a particular surface signature of a sub-surface vortex. 
To test the hypothesis we commence by identifying scars from the surface elevation alone and consider conditional and unconditional statistics of the flow beneath.

Heuristically we demand that the following are necessary conditions for a proper definition of a scar \emph{based on surface deformation only}:
\begin{enumerate}[align=left,leftmargin=1cm,itemindent=0pt,labelsep=0pt,labelwidth=2em]
    \item Scars are depressions in the surface; 
    \item Scars adhere to the surface for some time, long compared to noise imprints on the surface.
    \item Scars are elongated structures, i.e., they have areas much smaller than the smallest circle which circumscribes them.
    \item Scars are advected approximately with their surrounding flow.
\end{enumerate}
The final point is included to distinguish scars from propagating or standing surface waves. It is not of practical interest for the data we consider here since we do not have waves present, but becomes relevant in a field setting, and is included for completeness. Surface features will not in general move with precisely the mean flow since vortices also propel themselves. In a spectral sense, imprints of turbulence on a mean flow will produce broadband signals approximately on the advection line of the flow, not on the wave dispersion curve \citep{luo23, bullee24}.

We implement detection of scars by adapting the method of dimple detection using wavelets presented by \cite{babiker2023}. The conditions are met by: (i) using so-called `Mexican hat' wavelets; (ii) filtering out structures with shorter lifetime than five recorded frames in the simulations (approximatly $0.06 \Tinf$ for case 1); (iii) filtering out structures based on their eccentricity (an alternative to eccentricity is the roundness, $4\pi \times$Area$/($Perimeter$)^2$, which gives essentially identical results). As the detected structures change shape over time, we use running averages with a window size of five recorded frames to compute the eccentricity. 
Structures detected by the wavelets which have averaged eccentricity $<0.85$ are discarded, hence, we have reversed the condition that \cite{babiker2023} used to detect the near-circular dimples.
    
Consider the snapshot of the surface elevation in a region where the method above identifies a scar, in figure 
\ref{fig:scar}(a).
By visual inspection, a scar can be discerned as an elongated depression on the surface next to a near-circular dimple; these match the features detected with the method described above as shown in figure~\ref{fig:scar}(b).
Applying the $\lambda_2$ criterion on the flow below the scar brings a horizontal vortex tube aligned with the scar to attention (figure \ref{fig:scar}c). This is an example of the hypothesized `scar/submerged vortex'-coupling, a phenomenon we investigate statistically in the \S\,\ref{sec:probScars}. The well-known surface-attached vortex terminating in a surface dimple is also shown. The mean orientation of these vortices is roughly parallel and orthogonal to the surface, respectively, a point we explore in detail in \S\,\ref{subsec:inclAngles}.

\subsection{Conditional vortex probabilities beneath dimples} \label{subsec:vortProb}

We begin by considering surface-attached vortices, adopting a simple method: We numerically estimate dependent probabilities that a point $\rz=(x,y,\barz)$ lie inside a vortex given that the corresponding point $\reta=(x,y,-\eta(x,y))$ directly above lies within a (true) dimple. We denote these conditional probabilities $\fVetaz=P[V(\rz)|V(\reta)]$, with:
\begin{itemize}[align=left,leftmargin=1cm,itemindent=0pt,labelsep=0pt,labelwidth=2em]
    \item $V(\rz)$: The event of $\rz$ lying inside a vortex.%
    \item $V(\reta)$: The event of $\reta$ lying inside a dimple at the surface.
\end{itemize}
Consequently, $\fVz=P[V(\rz)]$ denotes the (unconditional) probability of being inside a vortex at depth $\barz$ for any arbitrary horizontal position. Since we identify dimples from surface-attached vortices, $\lim_{\barz\to 0}\fVetaz=1$ by construction
(using dimples detected from the SOC gives a value slightly smaller than $1$ due to the occasional false positive identification, but otherwise changes none of the conclusions). The results for both conditional and unconditional probabilities are averaged over horizontal grid planes and over time. The depth measure $\barz$ is used to account for the surface-adhering grid (see \S\,\ref{subsec:velvort}).

\begin{figure}
    \centering
    \includegraphics[width=0.5\linewidth]{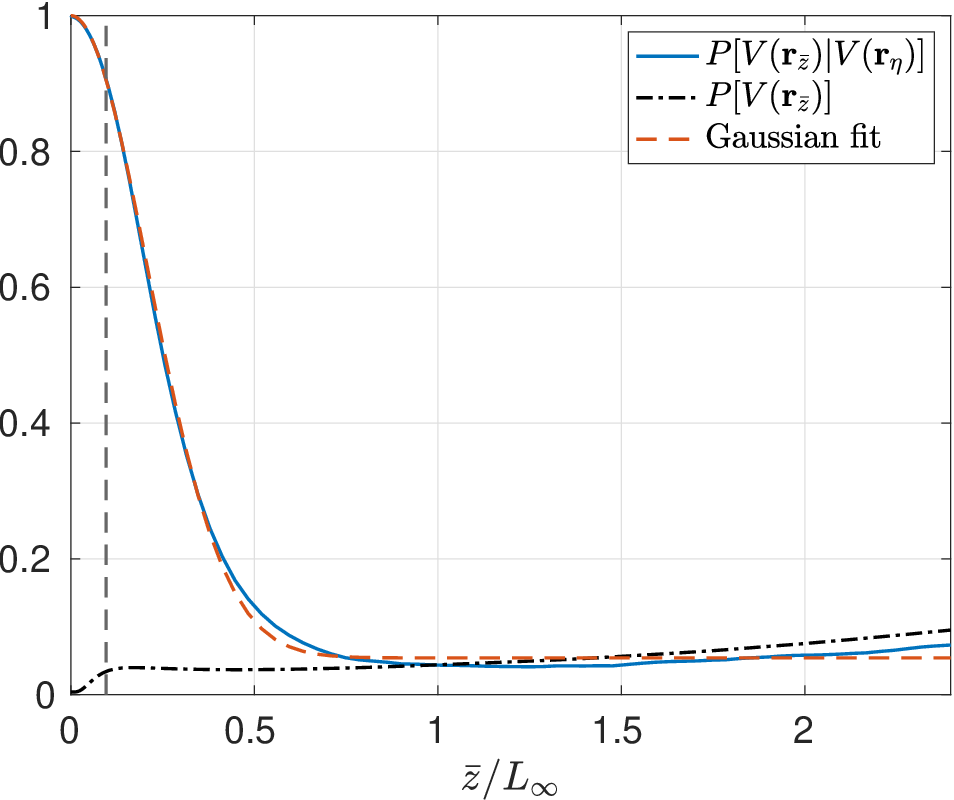}
    \caption{Conditional probability $P[V(\rz)|V(\reta)]$ of being inside a vortex, given that 
    the horizontal position lies directly beneath the centroid of the free-surface cross-section of a surface-attached vortex.
    Unconditional probability $P[V(\rz)]$ (black, dash-dotted line) and Gaussian fit (red, dashed line) for reference. The vertical dashed line marks the limit of the viscous boundary layer.}%
    \label{fig:probAllCase1a}%
\end{figure}

We plot these probabilities in figure \ref{fig:probAllCase1a} where each dimple is given the same weight regardless of its size and strength, i.e., for conditional averages only points directly beneath the centroid of the surface region wherein $\lambda_2\leq\lamth$ are considered. 

Figure \ref{fig:probAllCase1a} depicts the results for case 1, with conditional averages beneath vortex surface-centroids shown as a solid line along with a fitted Gaussian function for reference (dashed line), and unconditional probabilities (dash-dot line). The Gaussian fit closely follows the computed curve of conditional probability, particularly in the vicinity of the free surface. Moreover, the dependent probabilities $\fVetaz$ approach $\fVz$ in the outer part of the blockage layer ($z \gtrsim L_\infty$) where the influence of the free surface becomes negligible. Note, however, that $\fVetaz$ appears to dip below $\fVz$ rather than directly overlap with it. We return to this when discussing the equivalent probabilities for scars below. 

While the conditional probability $\fVeta(\eta)=1$ at the surface by construction, the unconditional probability $\fVz$ is nearly indistinguishable from zero at $\barz\to 0$ in figure \ref{fig:probAllCase1a}. This reflects how only a small fraction of near-surface vortices attach to the surface while the majority do not penetrate the viscous layer.

\begin{figure}%
    \centering
    \begin{overpic}[width=\textwidth,trim={1.4cm 1.4cm 3.2cm 0cm}, clip,grid=off,unit=1bp,tics=5]{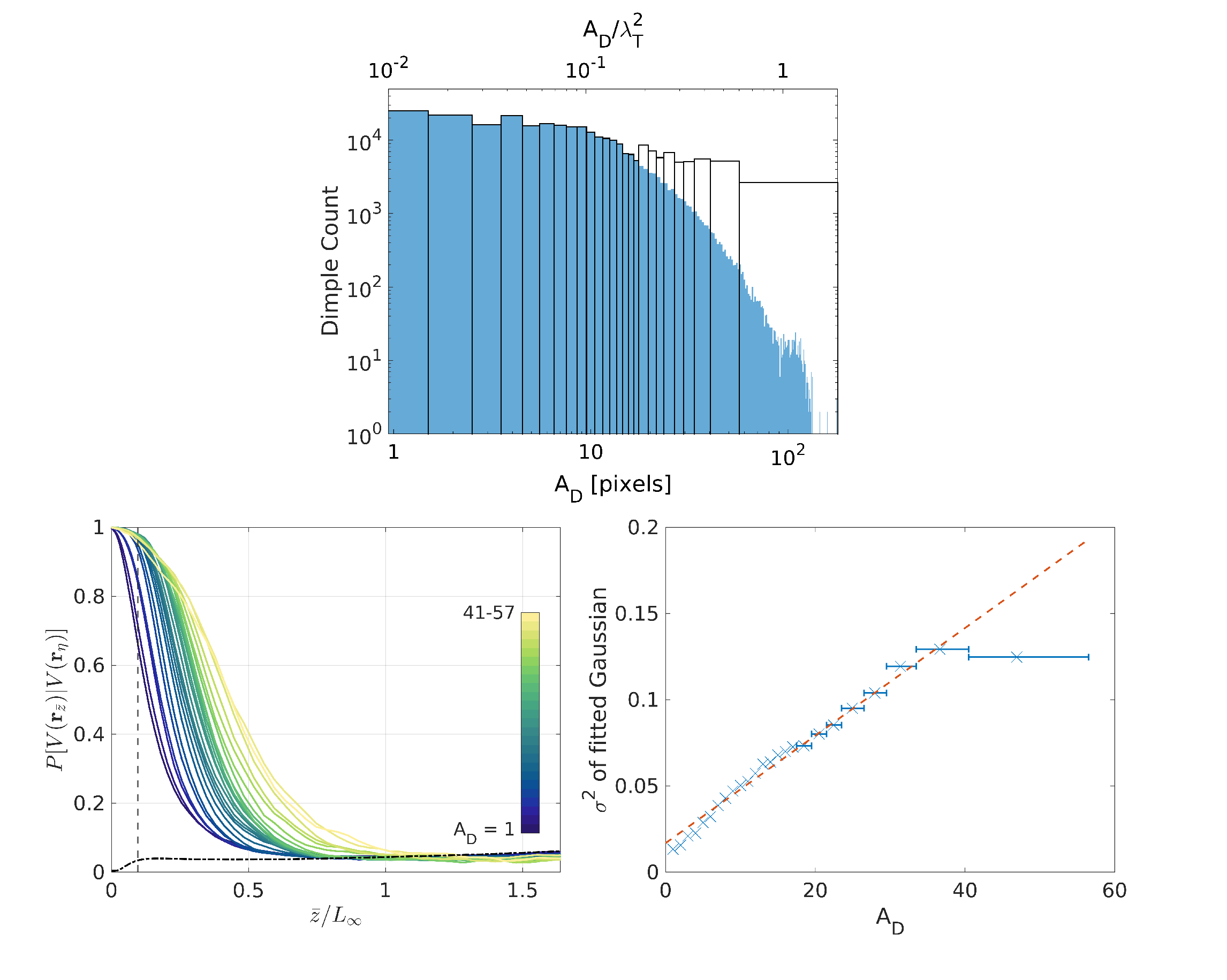}
    \put (69,75) {(a)}
    \put (43,33) {(b)}
    \put (95,33) {(c)}
    \end{overpic}
    \caption{(a) Distribution of vortex count by dimple area ($A_D$), the latter measured as the area of the cross-section of a surface-attached vortex at $\barz=0$ and given in number of pixels and as scaled by Taylor microscale squared (a single pixel has an area of approximately $10^{-2} \lambda_T^2$). Blue bars represent the vortex count per area bin, black rectangles delimit the bins so that each one, except for the rightmost one, contains at least $5000$ dimple counts. (b) Curves for conditional probability $P[V(\rz)|V(\reta)]$. 
    Each curve represents one bin from (a) coloured by increasing dimple area. The vertical dashed line marks the limit of the viscous boundary layer. The dash-dotted black line is the unconditional probability, $P[V(\rz)]$. (c) Variance ($\sigma^2$) of approximated Gaussian curves of conditional probability data in (b) by the weighted average of bin area (blue x-markers) and linear fit using $l_1$-norm (dashed red line). Horizontal error bars indicate the range of dimple area covered by bins that span multiple sizes.}
    \label{fig:vortSizeProb}%
\end{figure}

The computations of conditional probabilities are repeated, considering the dependency on vortex surface area (or more precisely: dimple area) by binning surface cross-sections of vortex tubes identified by the $\lambda_2$ criterion. 
Results for the conditional probabilities along with binning properties are shown in figure \ref{fig:vortSizeProb}. 

In the computation of size distribution and binning, vortex sizes are computed and counted for each time step, meaning that vortices that persist over multiple time steps are counted once for each time step. Although no weighting is performed with respect to dimple size or depth, larger vortices tend to have longer lifetimes and thus produce more counts. Also, one and the same vortex can be counted into different bins, as its surface cross-sectional area changes over time. The distribution of vortex count by size (blue bars in figure \ref{fig:vortSizeProb}a) is roughly exponential for vortices smaller than about $80$ pixels and levels off for the infrequent larger vortices. This could be an indication that a critical size exists above which the trend changes, although the comparatively small number of large vortices makes such conclusions tentative. 

To get statistically significant results when conditional probabilities are computed for each size span of vortices, we require each bin to contain at least $5000$ vortex counts, i.e., for each plotted curve of conditional probabilities in figure \ref{fig:vortSizeProb}(b). Where bins contain more than a single area, this is indicated with a black box in figure \ref{fig:vortSizeProb}(a). 
This results in a total of $25$ bins and ensures that bin averages are reasonably well converged. 
The rightmost bin (spanning vortices with area $57$--$178$ pixels) counts only $\approx 2650$ vortices and is therefore not included in figure \ref{fig:vortSizeProb}(b). 

Figure \ref{fig:vortSizeProb}(b) depicts the vortex size-dependent conditional probabilities as a function of depth $\barz$. Considering single curves, we notice that all the probability curves are near-Gaussian, like the probabilities averaged over all vortex sizes in figure \ref{fig:probAllCase1a}, while remarkably, the variance and probability, plotted in figure \ref{fig:vortSizeProb}(c), increase linearly 
with respect to the vortex size (except for the outlier for the bin with the largest vortices). 
All vortices greater than approximately $6$ pixels extend well beyond the viscous layer (probability $>95\%$), and even among the single-pixel vortices, $66\%$ remain present directly beneath the (tiny) dimple. 
Larger vortices penetrate much further, with the largest extending vertically into the lower part of the blockage layer. Nevertheless, all size-dependent conditional probability curves are below $0.38$ when measured at depth $\barz \geq 0.5 L_\infty$, and fall off to the average volume fraction of vortices within the blockage layer. 

\subsection{Conditional vortex probabilities beneath scars} \label{sec:probScars}

We next compute probabilities as described in \S\,\ref{subsec:vortProb}, but now conditional of $\reta$ lying inside an area of the surface identified as a scar with the SOC implementation in \S\,\ref{sec:scarIdent}. We calculate the conditional probability $\fSetaz=P[V(\rz)|S(\reta)]$, where $S(\reta)$ is the event of $\reta$ being inside a scar at the surface. Note that a slight modification is necessary to compute the conditional probabilities of the scars as compared to those of vortices: Since scars are non-circular, using their centroids as measurement points may entail considering points lying outside the scars themselves for the computations of conditional probabilities; this happens for approximately $6\%$ of all scars for case 1 when centroids are used without modification. A more fitting measurement point for a scar is a combination of the centroid and its approximated centreline. The latter is computed by \textit{skeletonisation}. This process reduces a two-dimensional pattern to a centreline of single-pixel thickness while preserving its topology (we use the function \textit{bwskel} in MATLAB's Image Processing Toolbox for this purpose. See documentation \cite{MATLAB} and references therein). After skeletonisation, the measurement point is set to the point on the centreline that has the shortest Euclidean distance to the centroid. This final step serves to avoid overemphasizing long, thin tails of scars, as may happen if the midpoint of the centreline is selected directly.

\begin{figure}%
    \centering
    \includegraphics[width=0.5\linewidth]{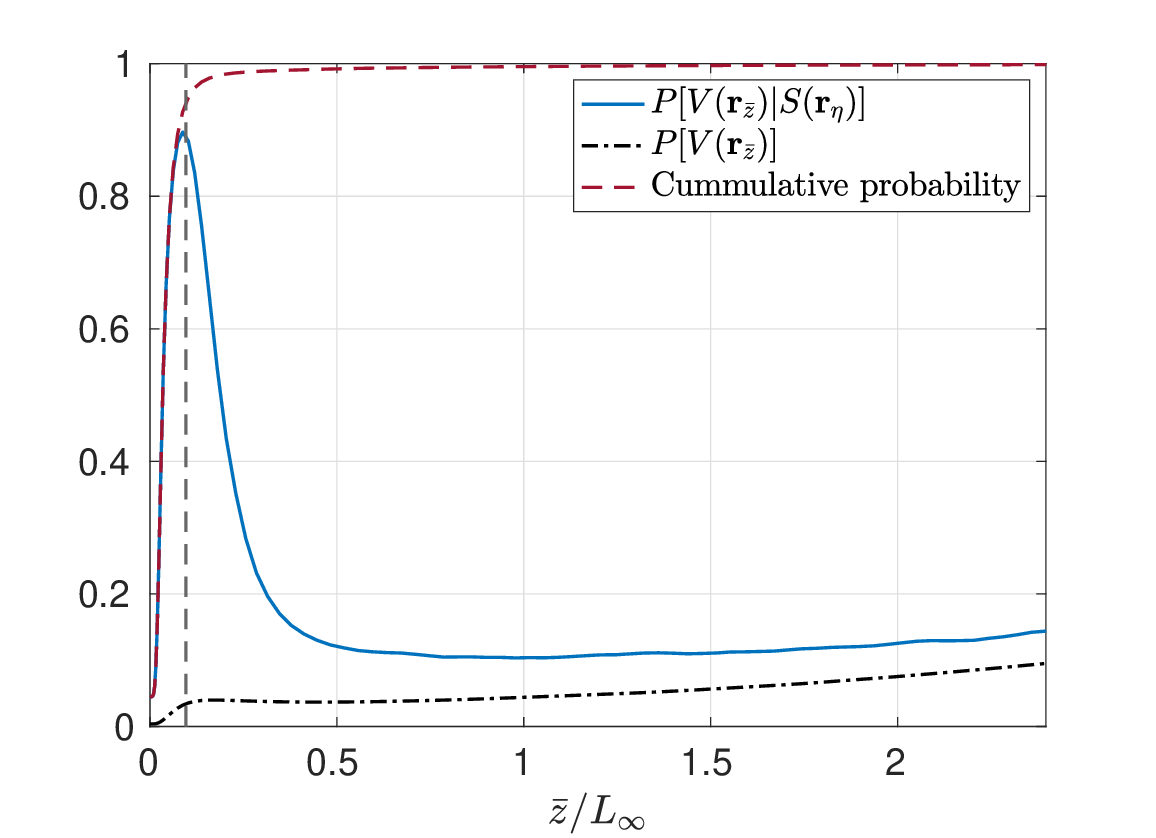}
    \caption{Conditional probability $P[V(\rz)|S(\reta)]$ of $\rz$ being inside a vortex, given that there is a scar on the surface directly above (blue, solid line), the corresponding cumulative probability (red, dashed line), and the unconditional probability $P[V(\rz)]$ (black, dash-dotted line), for reference. 
    The vertical dashed line marks the limit of the viscous boundary layer.}%
    \label{fig:probAllCase1aScar}%
\end{figure}

Figure \ref{fig:probAllCase1aScar} depicts the conditional vortex probabilities beneath scars for case 1. 
One immediately observes that the conditional probability for vortices beneath scars increases rapidly through the viscous layer, with maximum probability at the limit of this layer. In other words, virtually all structures that were identified as scars from the set of SOC have a vortex directly beneath them situated approximately at $z=L_\nu$. Correspondingly, by considering the cumulative probability, i.e., the probability that there is at least one point within a vortex along the vertical line from the surface down to a level $\barz$, we see that it has reached more than $90\%$ by $\barz=\Ln$. Starting from an arbitrary point, the same probability is only 3.4\%. On this evidence, we can safely conclude that a scar, as identified with the set of SOC in \S\,\ref{sec:scarIdent}, is a signature of a vortex situated around $\barz\approx \Ln$, especially since our scar identification implementation includes a nonzero number of false positives.

Another observation from figure \ref{fig:probAllCase1aScar}
is that unlike $\fVetaz$, $\fSetaz$ does not approach $\fVz$ inside the depth range plotted here. This is not surprising given the intermittency of the turbulence and the distribution of scars: Scars are numerous only during strong turbulent (upwelling) events (see Supplementary Material for videos depicting the evolution of scars and dimples at the surface). 
Thus, the conditional statistics for scars are biased towards times with increased vortex concentration in the upper bulk. The amount of dimples present in the flow is also connected to upwelling events. Yet, unlike scars, which spatially and temporally appear in the immediate vicinity of the upwelling structures, dimples appear as as upwelling boils mature and die out (with an associated lowering of vortex concentration), and are subsequently spread out and advected away from the upwelling cores. In fact, a weak opposite effect on conditional probabilities for dimples can be seen in figure \ref{fig:probAllCase1a}: $\fVetaz$ approaches $\fVz$ yet lies slightly below it in the bulk. These trends for $\fVetaz$ and $\fSetaz$ are even more prominent for the low Reynolds number cases (see \S\,\ref{sec:ReWe_effects}), where lower turbulence intensity overall increases the spatio--temporal dependencies of the instantaneous conditional probabilities.

\begin{figure}%
    \centering
   \begin{overpic}[width=\textwidth,trim={1.4cm 1.4cm 3.2cm 0cm}, clip,grid=off,unit=1bp,tics=5]{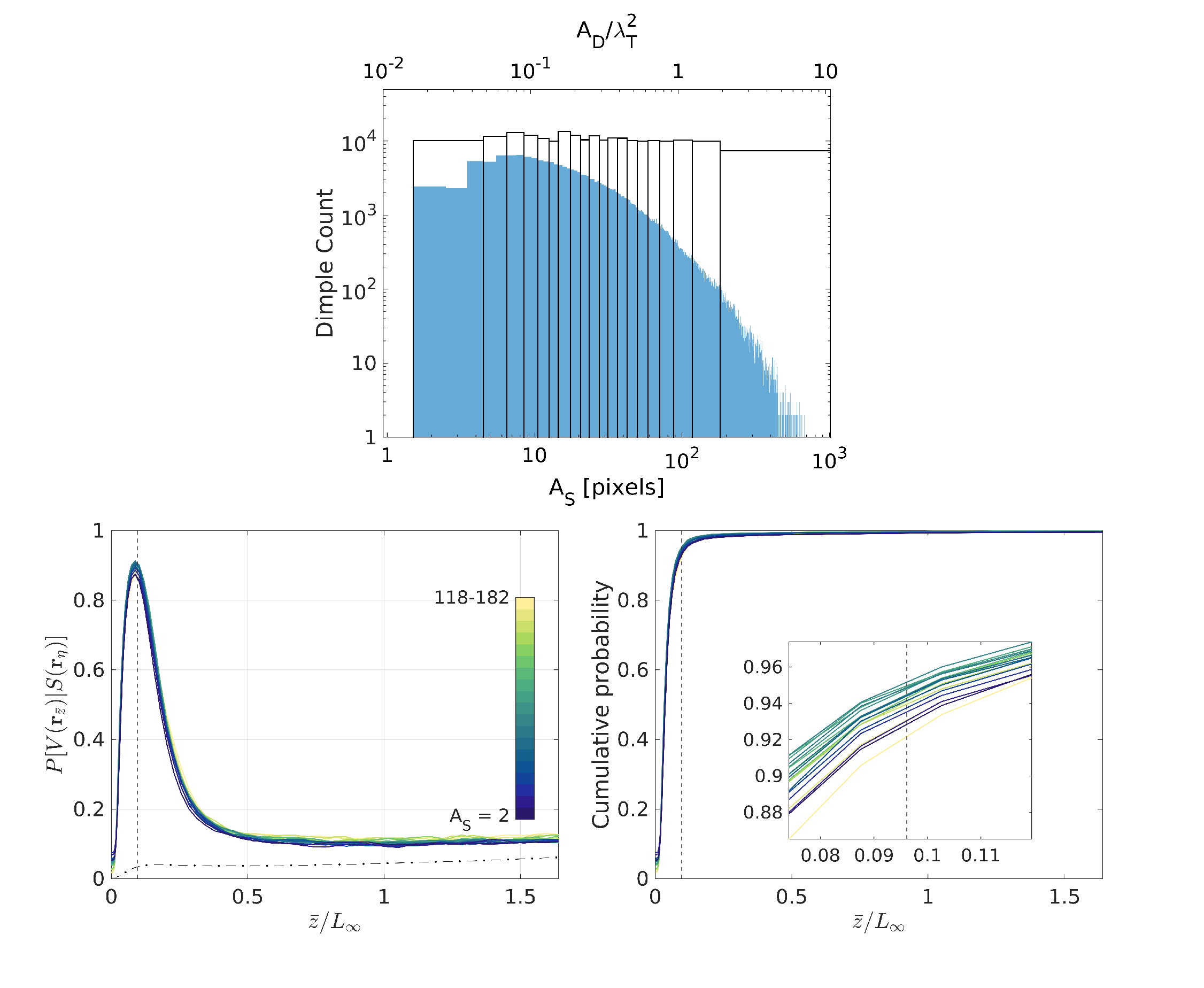
   }
    \put (69,75) {(a)}
    \put (43,34) {(b)}
    \put (95,34) {(c)}
    \end{overpic}
    \caption{(a) Distribution of scar count by scar area ($A_S$), where the latter is the surface area covered by a detected scar in number of pixels and as scaled by Taylor microscale. Blue bars represent the scar count for each pixel size. Black rectangles denote the bins used for the computation of conditional probabilities. 
    (b) Curves for conditional probability $P[V(\rz)|S(\reta)]$ of being inside a vortex, given that there is a scar at the surface at the same $x,y$-position, sorted by scar area so that each curve represents one bin in (a). The vertical dashed line marks the limit of the viscous boundary layer. The dash-dotted black line is the unconditional probability, $P[V(\rz)]$. 
    (c) The corresponding cumulative probabilities, by scar area. Inset: Zoom in on the region where curves cross from the viscous layer to the blockage layer. 
    }%
    \label{fig:scarSizeProb}%
\end{figure}
By the same binning and averaging that we presented for dimples above, results for conditional and cumulative vortex probabilities beneath scars are computed for different scar sizes and shown in figure \ref{fig:scarSizeProb}, along with binning details. 
Since the scars in our simulations are less numerous and longer lasting (on average) than the dimples, we require more instances in each bin for scars to avoid the danger of single bins containing only a small amount of long-lasting scars (recall that each count denotes an imprint at a single time step). Therefore, each bin contains at least $10000$ counts to ensure statistical convergence, except for the one containing the largest scars, which does not reach the minimum number and is excluded from the analysis. 

The difference between figures \ref{fig:vortSizeProb}  and \ref{fig:scarSizeProb} is striking. Whereas the probability of finding vortices a finite distance beneath a dimple shows a strong dependence on dimple area, the curves in figure \ref{fig:scarSizeProb}(b,c) collapse near perfectly and show no dependence on scar size.  
The results very clearly support the conclusion that below every scar there is a vortex situated near the lower edge of the viscous layer, $\barz\approx\Ln$, regardless of the size of the scar. Towards the formulation of a PCD we conclude that a true scar is an imprint of a sub-surface vortex which is not connected to the surface. Further statistical insight into these sub-scar vortices comes to light in \S\,\ref{subsec:inclAngles}.

\subsection{Reynolds number and Weber number effects on sub-feature statistics}\label{sec:ReWe_effects}

\begin{figure}%
    \centering
    \begin{overpic}[width=\textwidth,grid=off,tics=2]{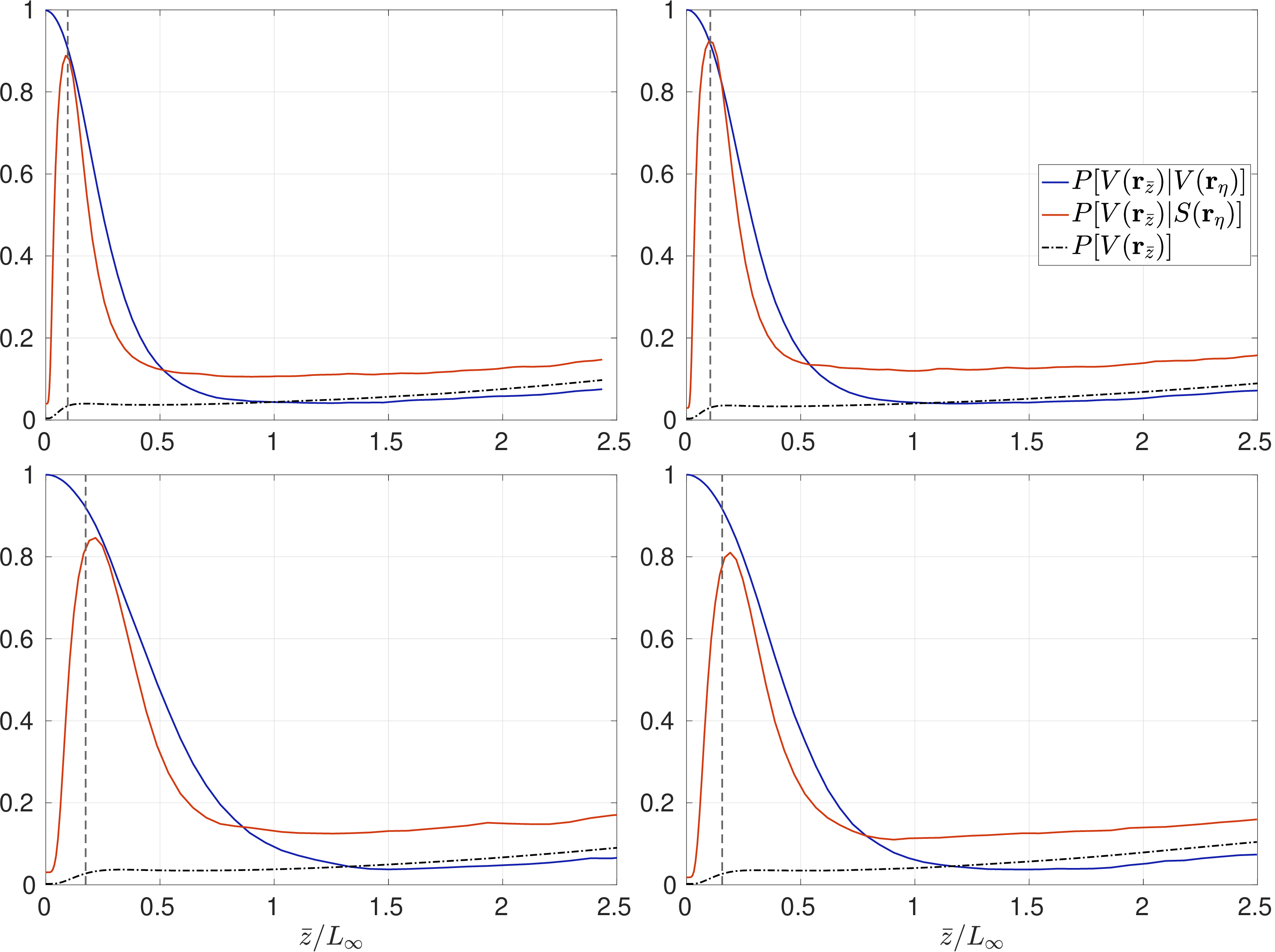}
    \put (41,69) {(a)}
    \put (91,69) {(b)}
    \put (41,32) {(c)}
    \put (91,32) {(d)}
    \end{overpic}
    \caption{Conditional probability for dimples, $P[V(\rz)|V(\reta)]$ (blue, solid line) and scars, $P[V(\rz)|S(\reta)]$ (red, solid line), for cases 1, 2, 3 and 4 (panels a--d, respectively). In each panel, the vertical dashed line marks the limit of the viscous boundary layer and the dash-dotted black line is the unconditional probability, $P[V(\rz)]$.}
    \label{fig:ReWeDep}%
\end{figure}

Consider figure \ref{fig:ReWeDep}, depicting the conditional probabilities $\fVetaz$ and $\fSetaz$ for all four cases detailed in table \ref{tab:flowProp}. As for case 1, also for cases 2--4 $\fVetaz$ falls off approximately as a Gaussian through the blockage layer, whereas $\fSetaz$ increases rapidly in the vicinity of the surface, with a maximum near $\barz = L_\nu$. 

Comparing the results for different Weber numbers reveals a negligible effect of surface tension on the conditional probabilities (with the caveat that we have considered only two values of $We$). The Reynolds number, on the other hand, does affect the statistical depth of the sub-surface vortices for both dimples and scars. 

\begin{figure}%
    \centering
    \includegraphics[width=0.7\linewidth]{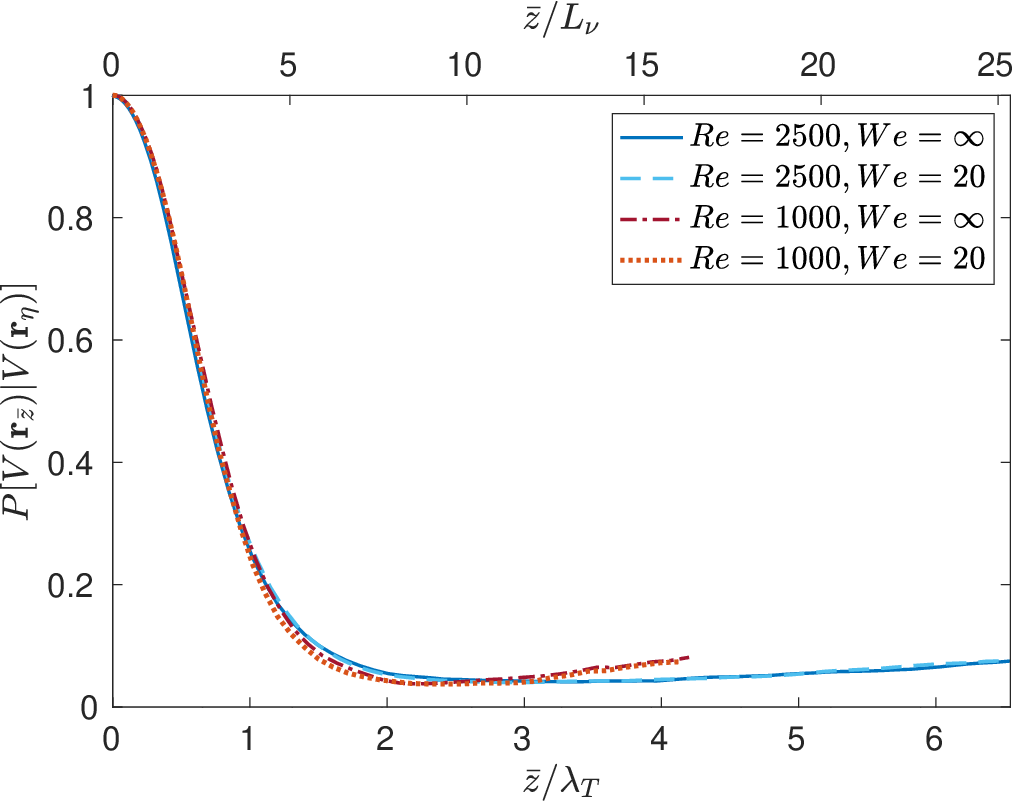}
    \caption{Conditional probabilities for dimples for different Reynolds numbers and Weber numbers. Scaling by Taylor microscale ($\lambda_T$) or viscous layer thickness ($L_\nu$).}%
    \label{fig:ReWeDepScaled}%
\end{figure}

A different scaling of the abscissa points in the direction of what drives the Reynolds number effect. Observe in figure \ref{fig:ReWeDepScaled}, where the conditional probabilities $\fVetaz$ are plotted as a function of depth scaled by $L_\nu$ (or $\lambda_T$, discussed below) rather than $\Linf$, that all curves for $P[V(\rz)|V(\reta),\mathrm{Re},\mathrm{We}]$ collapse onto one curve $P[V(\rz/L_\nu)|V(\reta)]$ in the vicinity of the surface. The same effect is evident for scars in figure \ref{fig:ReWeDep}, by the peak in $\fSetaz$ around $\barz = \Ln$ observed for all Reynolds numbers.
Further from the surface, some discrepancy between the results for $Re= 2500$ and $Re= 1000$ can be observed. This is due to the rapid decay of the turbulence rising from the deeper bulk for the lower Reynolds number cases. 
 
Recalling that the thickness is computed by $L_\nu \approx 2 Re_\infty^{-1/2} L_\infty$ and inserting the definitions of $\Linf$ and $Re_\infty$
(see \S\,\ref{subsec:flowProp}) yields:
\begin{align}
    L_\nu & \approx  
    \frac{\lambda_T}{\sqrt{15}}\, \approx 0.26\,\lambda_T ,
\end{align}
i.e., a purely Taylor microscale-dependent number. Hence our study supports the conclusion that the vertical reach of a surface-attached vortex beneath a dimple depends on the Taylor microscale of the turbulence in the bulk only, independent of surface tension. In terms of conditional probability, it may be expressed as 
\begin{equation}
    \fVeta(\barz, Re,We) \to \fVeta(\barz/\lambda_T).
\end{equation}
Note, however, that this conclusion is shown for a small sample of $Re$, to hold within a limited range of Reynolds and Weber numbers. Further exploration is called for to establish the limits of applicability.

\subsection{Inclination angles for vortices below scars and dimples} \label{subsec:inclAngles}

The conditional probabilities for finding vortices directly beneath dimples and scars give insight into the prevalence of vortices, but not their properties. Since turbulence near the surface is highly anisotropic, a key property of a vortex is its orientation. We therefore now consider the inclination angles of vortices below said patterns. 

Vortex inclination angles $\theta_{zx}$ are computed from \eqref{eq:thxz}, now considering only grid points that are covered (from above) by a scar or a dimple, i.e., from components of the vorticity field $\bom(\hx,\hy,\barz,t)$ where $\hat{x}, \hat{y}$ denote the time-dependent subset of $x,y$-coordinates for which $V(\reta) = 1$ for dimples and $S(\reta) = 1$ for scars at time $t$. 

\begin{figure}%
    \centering
    \includegraphics[width=1\linewidth]{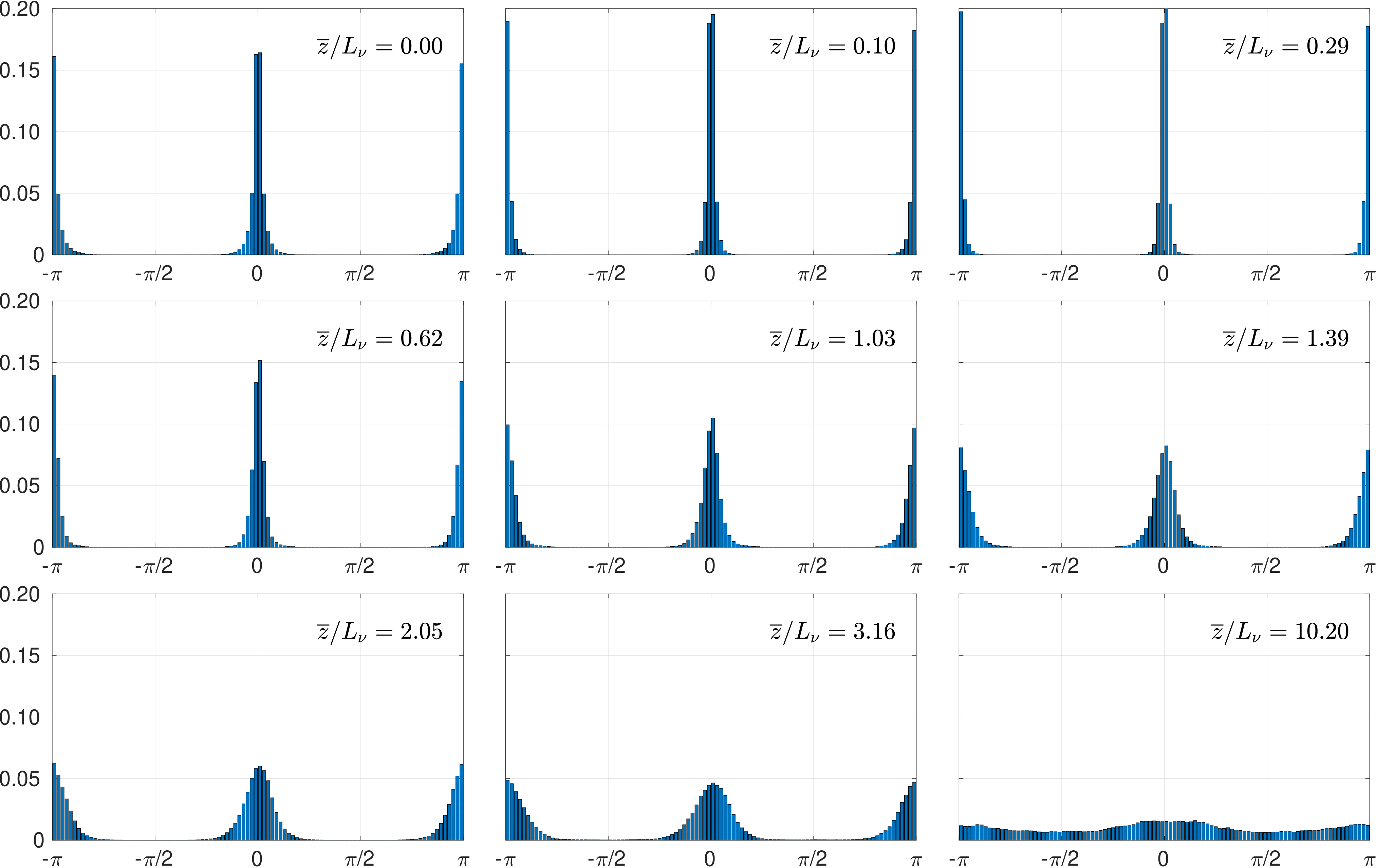}%
    \caption{Normalised histograms of weighted vortex inclination angles for case 1, at increasing average depths spanning the surface and blockage layers, including only regions below vortex dimples. All depths are given in multiples of the viscous surface layer thickness.}%
    \label{fig:inclAnglVort}
\end{figure}

\begin{figure}%
    \centering
    \includegraphics[width=\linewidth]{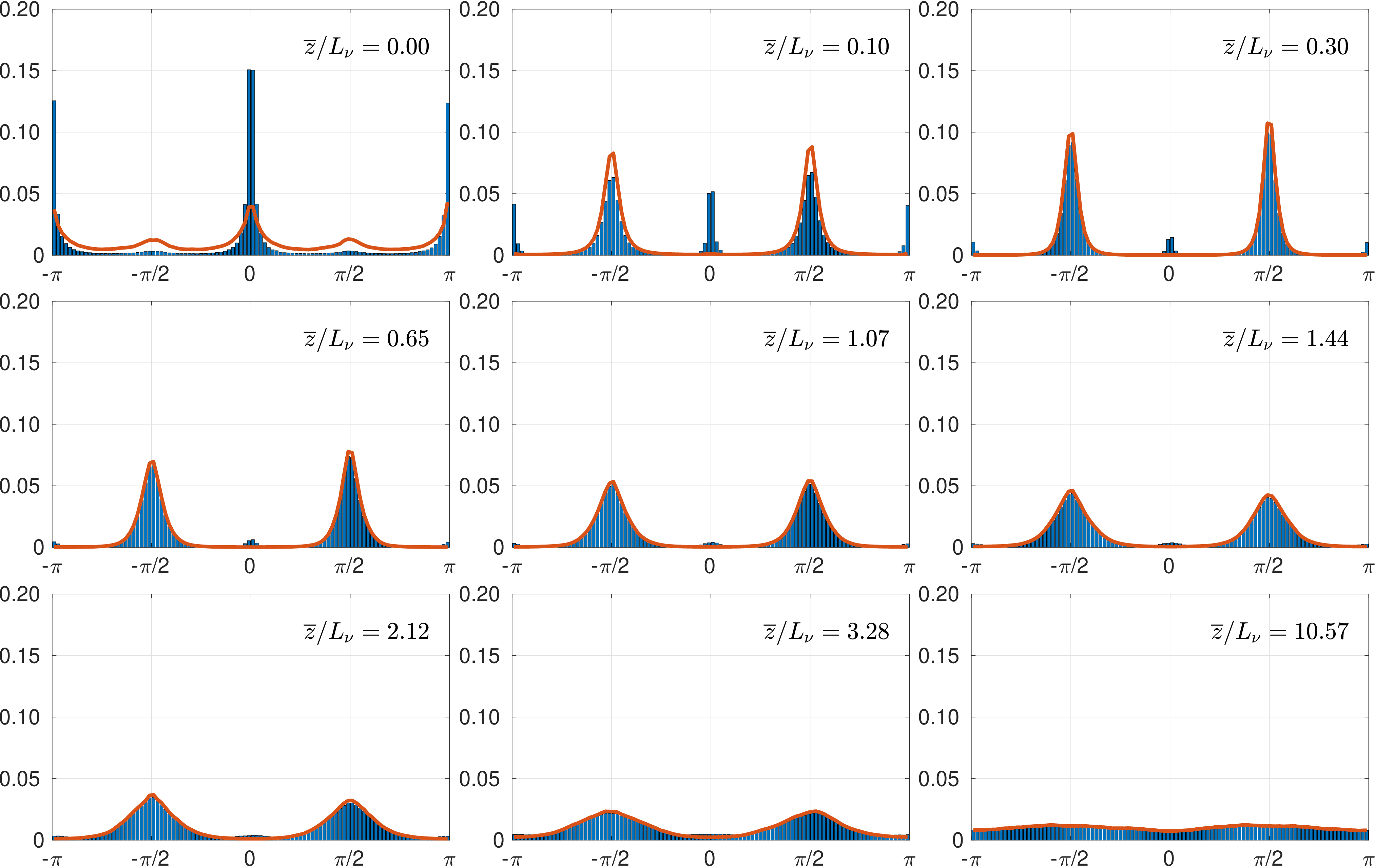}%
    \caption{
     The same as figure \ref{fig:inclAnglVort}, but now for regions beneath scars. The added red line represents results for scars that do not overlap with dimples.
    }%
    \label{fig:inclAnglScar}
\end{figure}

Figures \ref{fig:inclAnglVort} and \ref{fig:inclAnglScar} depict the resulting weighted inclination angle distributions at different depth levels beneath dimples and scars, respectively, for case 1. The most obvious conclusion from the figures is --- as predicted above --- that the vorticity is predominantly vertical beneath dimples and predominantly horizontal beneath scars. 

The trend from the results of the inclination angles is clear when compared to the inclination angles of the vorticity vector in the flow in general (figure \ref{fig:inclAngl}). Recall that we had a transition from one preferred vortex inclination angle (vertical; $\theta_{zx} = 0$ and $\pm \pi$) to another (horizontal; $\theta_{zx} = \pm \pi/2$) when starting at the free surface and moving through the viscous layer. More precisely, the unconditional distribution of vorticity inclination shows that inside the viscous layer, the vorticity field is either near-horizontal or near-vertical with only a small fraction of points where $|\bom|$ is significant having intermediate orientations. At the surface all vorticity is vertical (or, more properly, surface-orthogonal) due to the viscous effect, gradually shifting to being dominated by horizontal orientation for $\barz\sim L_\nu$, an effect of blocking. The vorticity field remains anisotropic throughout the blockage layer, with vertical and horizontal orientations being more probable.

As observed in figures \ref{fig:inclAnglVort} and \ref{fig:inclAnglScar}  the picture is very different for dimples and scars. The vorticity below dimples at the surface is vertical by definition, yet more interestingly, the preference for horizontal orientation is entirely absent below the dimples. 
The statistical indication is that surface-attached vortices extend directly downwards, typically penetrating deep into the blockage layer before curving away, and deflecting away any horizontal vortices in the vicinity from entering the space beneath them.

A consideration of vorticity inclination below scars reveals the opposite trend. There is a strong bias towards horizontal orientation throughout the viscous layer and horizontal vortex inclination remains the preferred inclination well into the blockage layer. In comparison with the inclination angle results for the flow in general, shown in figure \ref{fig:inclAngl}, and for the flow below dimples, we notice that vertical components of the vorticity are small directly below the surface and all but extinguished below scars as early as at depth $\barz / L_\nu = 0.3$. 

The blue bars in figure \ref{fig:inclAnglScar} display a significant peak around the vertical orientations for the two topmost depths, at and very close beneath the surface. This initially puzzling fact is readily explained, however, from imperfections in our scar detection. Figure \ref{fig:scar} shows an example situation where a dimple lies close to a scar. Occasionally an area identified as a scar also contains a dimple. Moreover, although there is a high peak in the blue bars near $0$ and $\pm\pi$ for the two top depths, the vorticity itself is very small here so the impact of a very small number of dimples among the scars will be high.

After excluding the approximately $5\%$ scars (identified with the SOC) which contain a true dimple (identified with the $\lambda_2$ criterion), the red curves in figure \ref{fig:inclAnglScar} results. (A more sophisticated implementation of scar identification from the set of SOC would have mitigated this problem without reference to sub-surface flow --- we allow ourselves this pragmatic fix in order to demonstrate the physics involved.) But for the very surface itself, which is ambiguous, beneath the `corrected scars' the vertical vorticity preference is completely distinguished throughout the blocking layer, until the turbulence eventually becomes isotropic. Hence, we conclude towards the formulation of a PCD that the sub-surface vortex which imprints the surface as a scar is horizontally aligned.

Furthermore, considering the orientation of vortices below scars in light of the results in \S\,\ref{sec:probScars}, points to a cross-sectional length of $\mathcal{O}(L_\nu)$ of vortices located at the lower edge of the viscous layer below scars, regardless of the scar size. The longitudinal lengths of these vortices are, however, not clearly correlated to the scars: The surface imprint is very sensitive to the exact distance between vortex core and surface, and vortices winding slightly up and down in a serpentine manner may give rise to multiple scars when approaching the viscous boundary layer from below in multiple places along its length. Thus a long vortex can give rise to a much smaller scar, or even several shorter ones.

\section{Conclusions}

We have performed a statistical investigation of the vortex flow beneath two types of surface features: dimples and scars. The former has been much studied and is fairly well understood while the latter has received little attention in the past. We make use of data from four direct numerical simulation, with two different Reynolds numbers and two Weber numbers (with and without surface tension). As is well known the flow beneath the surface has distinct behaviour inside the viscous surface layer of width $\Ln$ equal to approximately a quarter of the Taylor microscale $\lambda_T$, and the blockage layer underneath which extends down to depth of approximately one integral scale of the bulk turbulence.

Our goal is to establish what statistical inferences can be drawn about sub-surface flow from observing the surface only, a question of significant practical importance. Towards this end we establish a pair of concepts which, when suitably formulated, prepares the ground for interpretation of observed surface features: surface-only criteria (SOC) which allow a type of imprint (here: dimples and scars) to be identified only from information about the free-surface elevation, and a pseudo-causal definition (PCD) which defines an observable surface feature of a type as the imprint of an identifiable sub-surface flow structure.

For dimples a SOC-PCD pair can be readily formulated from previous work: a set of SOC is that a dimple as a depression in the surface which is approximately circular and persists for a long time (compared to other relevant scales). Conversely, a `true dimple' can be defined as the imprint of a vertical vortex tube attached to the surface, appearing where the tube terminates (PCD). This SOC-PCD pair has been previously shown to be in excellent agreement (i.e., they identify the same areas of the surface as dimples).

For scars we here propose a corresponding set. SOC: a scar is a depression in the surface that is strongly elongated and lives for a long time; PCD: a `true scar' can be defined as being the imprint of a horizontal vortex beneath it, centred approximately at the lower edge of the surface viscous layer, a depth equal to approximately a quarter of the Taylor microscale. 

We present statistical evidence to demonstrate that this understanding is a successful description. First, we considered the conditional probability that a point in the flow belongs to a vortex core (according to the $\lambda_2$ criterion), given that the point lies directly beneath a dimple or scar. Then the corresponding statistics of the orientation of the vorticity field beneath the same features were found. We found thus that nearly all vortices extending downwards from dimples reach throughout the viscous sublayer and well into the blockage layer, and the largest vortices can reach all the way to the bottom of said layer where the flow is approximately isotropic. The vortex probability beneath dimples decreases with depth $\barz$ as a Gaussian with standard deviation increasing approximately linearly with the area of the dimple. For varying Reynolds and Weber numbers the dependence of the conditional probability on $\barz$ is a function of $\barz/\lambda_T$ and independent of surface tension (for the range of values considered here). Moreover, we found that the horizontal vortex beneath scars is invariably centred at the lower edge of the surface viscous layer, independently of the size of the scar. Going downwards beneath the scar, the cumulative probability of having encountered a vortex increases from negligible to $>95\%$ (averaged over our different cases) in the range $0<\barz<1.5\Ln$. After removing a small amount of regions identified as scars which also contain a surface-attached vortex (when the two are so close that our simple detection method mistakes it for a part of the scar), we observed that there is practically no vertically oriented vorticity at all beneath scars within the viscous and upper half of the blocking layer, all having been deflected by the horizontal vortex situated there. A preference for horizontal vorticity persists beneath scars throughout the blockage layer. 

The statistical relations we have found between surface shape and sub-surface vortex structures not only support the suggested SOC-PCD pairs for dimples and scars, but also serve to elucidate further the nature of the coherent vortex structures in the vicinity of the turbulent free-surface flow, as well as their connection to imprints on the surface. While it is well known that the turbulence changes gradually through the blockage layer from being essentially isotropic outside of it to becoming highly anisotropic as the viscous layer is approached, our statistical study of vorticity inclination angles --- and by inference the orientation of vortex tubes --- offers a much more detailed and quantitative description. 
Surface-attached vortices, already well-known to appear beneath dimples, are found to penetrate deep into the blocking layer, the largest reaching all the way to $\barz\sim \Linf$; this is seen by how the vorticity which is necessarily normal to the surface at $\barz=0$, retains a strong vertical preference throughout the blocking layer, with virtually no horizontal vorticity present in its upper half. 
Correspondingly, horizontally oriented vortex tubes associated with scars cluster at the edge of the viscous boundary layer, around $\barz\approx L_\nu$ with remarkable consistency rather than penetrating it. 
As such, the viscous boundary layer acts as a `soft boundary', its lower edge demarking the depth where vortex tubes are forced to either penetrate the boundary layer and attach to the surface, or otherwise arrange parallel to it. 

We are thus able to conclude that scars on the surface can be used to identify the position of strong, horizontally oriented vortex tubes, similar to how dimples show the location of vertical ones. One might say that 
dimples \textit{directly} disclose surface-attached vortical structures and scars \textit{indirectly} disclose sub-surface vortical structures, in the sense that there can be no surface-parallel vorticity at the surface itself. 

\backsection[Acknowledgements]{We have benefited greatly from discussions with the members of the free-surface flows research group at NTNU.
The research of J.R.A. and S.\AA.E. was funded by the Research Council of Norway (iMod, 325114) and S.\AA.E. also by the European Union (WaTurSheD, ERC grant 101045299). Views and opinions expressed are however those of the authors only and do not necessarily reflect those of the European Union or the European Research Council. Neither the European Union nor the granting authority can be held responsible for them. The research of A.X.\ and L.S.\ was funded by the Office of Naval Research.}

\backsection[Declaration of interests]{The authors report no conflict of interest.}

\appendix

\section{Appropriate choice of $\boldsymbol{\lambda}_{\boldsymbol{2},\mathrm{\textbf{th}}}$}
\label{app:lam2threshold}

The value of $\lamth$ influences the probability that a given point lies 
inside a vortex, so care is needed when specifying this threshold.
The original suggestion by \citet{jeongIdentificationVortex1995} to use $\lamth=0$ makes for an oversensitive detection method with many false positives, while with $|\lamth|$ too large the method fails to detect the weaker vortices. 
There is no established standard for determining $\lamth$ in free-surface turbulence, and the values deemed appropriate vary among authors \citep[e.g.,][]{nagaosa03,khakpour12}.

A common approach is to normalise $\lambda_2$ using a characteristic frequency, $\OmT$, based on the inverse time scale of the vortex structures in the flow. \cite{nagaosa1999} and \cite{khakpour12} found this frequency by taking the ratio of the frictional velocities and the depths of their channels, while \cite{schram04} used the incoming velocity and step height in their flow case of a backward-facing step. In our case, we choose $\urms$ and the Taylor microscale to define the characteristic frequency as $\OmT^2 \equiv (\urms/\Taylor)^2=2\overline{s_{ij}s_{ji}}/15$, where we assume that an appropriate threshold should scale with this frequency, $\lamth \propto \OmT^2$.

\begin{figure}%
    \centering
    \includegraphics[width=\linewidth, trim={3cm 0cm 3cm 1cm}, clip,]{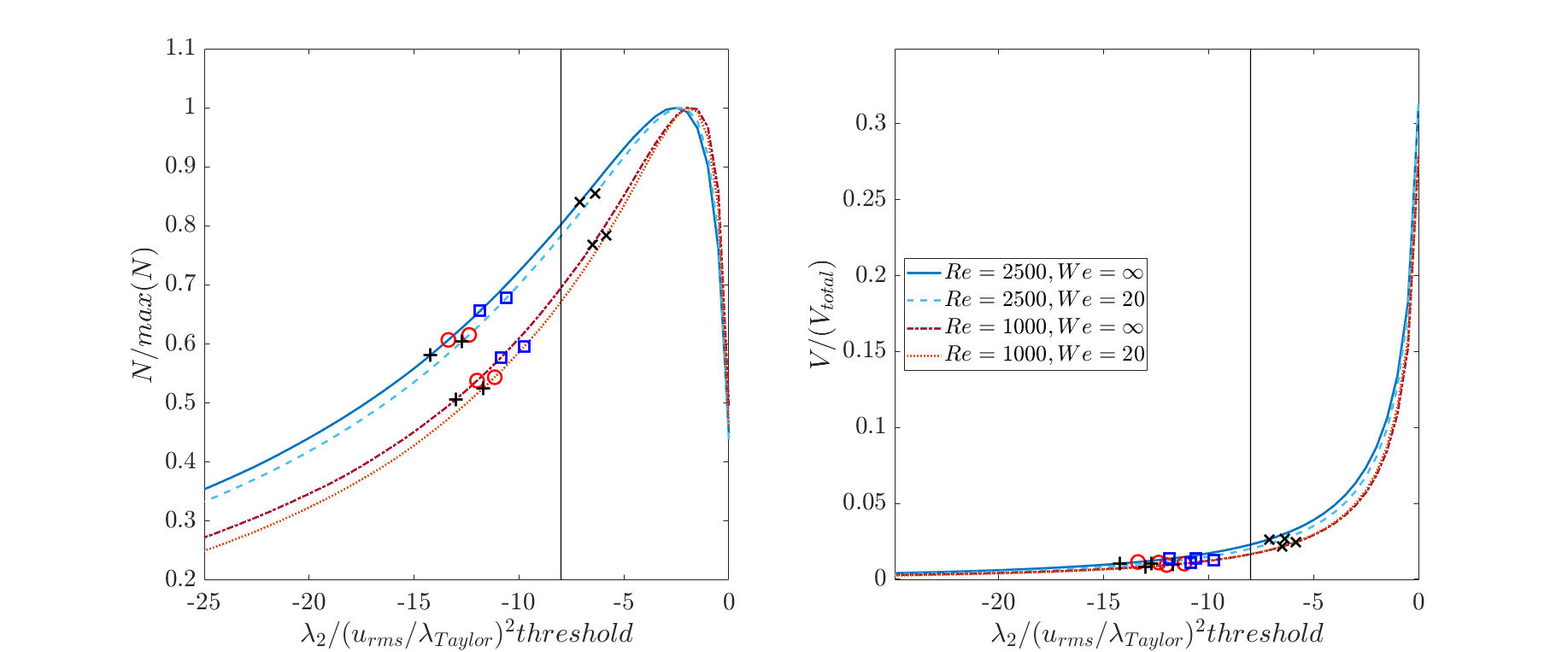}
    \caption{The effect of varying $\lamth$ on the number of retained vortex structures normalised by the peak number (a), and the volume of retained vortex structures normalised by the total volume of the free region (b). Also included are markers when choosing other methods to find these thresholds: red circles: \cite{schram04}, black squares: \cite{jeong1997}, crosses: lower threshold by \cite{khakpour12}, pluses: upper threshold by \cite{khakpour12}}%
    \label{fig:lambda2app}%
\end{figure}

Consider figure \ref{fig:lambda2app}, which depicts the normalised vortex count (figure \ref{fig:lambda2app}a) and the volume fraction (figure \ref{fig:lambda2app}b) for different $\lamth$ (normalized with the bulk properties) for the four flow cases we have detailed above. The results indicate that the scaling $\lamth\propto\OmT^2 $ and our selected threshold, $\lamth/\OmT^2=8$ (represented by a vertical black line in each panel),  are not unreasonable. The $\lambda_2$ threshold is
selected as a compromise where most small and short-lived vortices are detected, but mere fluctuation outliers are mostly discarded. Note in passing that this implies $\lamth\approx\overline{s_{ij}s_{ji}}$.  Thresholds used in other studies are indicated with markers. 

Alternatives to using the $\lambda_2$ criterion for vortex identification can be found in the literature. The method by \cite{zhou99} uses $\lamci$, the complex component of the eigenvalues of the velocity gradient tensor. Similarly to using the $\lambda_2$ criterion, the method requires a threshold for $\lamci$ to identify vortices. We set the threshold such that all points with $\lamci/\OmT > 3$ belong to a vortex core. Applying this to our data and comparing it to the vortex detection with the $\lambda_2$ criterion yields a spatial correlation of detected vortices of approximately 91\% for all four cases. This indicates that choosing one of these methods over the other will not have a significant impact on the analysis of our data sets.

\bibliographystyle{jfm}
\bibliography{vortex_corr}

\end{document}